\documentclass[11pt,onecolumn,draftcls]{IEEEtran}
\newtheorem{theorem}{Theorem}
\newtheorem{proposition}{Proposition}
\newtheorem{corollary}{Corollary}
\newtheorem{definition}{Definition}
\newtheorem{lemma}{Lemma}

\newcommand{\df}{\stackrel{\mbox{\scriptsize def}}{=}}
\newcommand{\ELS}{ELS}
\newcommand{\rk}{\mathrm{rk}}

\newcommand{\dr}{d_{\mbox{\tiny{R}}}}
\newcommand{\Kr}{K_{\mbox{\tiny{R}}}}

\newcommand{\vspan}[1]{\left< #1 \right>}

\usepackage{amsmath,amssymb,cite,subfigure}
\usepackage[dvips]{graphicx}

\begin{document}
\title{Properties of Rank Metric Codes}
\author{Maximilien Gadouleau and Zhiyuan Yan\\
Department of Electrical and Computer Engineering \\
Lehigh University, PA 18015, USA\\ E-mails: \{magc,
yan\}@lehigh.edu} \maketitle

\thispagestyle{empty}

\begin{abstract}
This paper investigates general properties of codes with the rank
metric. First, we investigate asymptotic packing properties of rank
metric codes. Then, we study sphere covering properties of rank
metric codes, derive bounds on their parameters, and investigate
their asymptotic covering properties. Finally, we establish several
identities that relate the rank weight distribution of a linear code
to that of its dual code. One of our identities is the counterpart
of the MacWilliams identity for the Hamming metric, and it has a
different form from the identity by Delsarte.
\end{abstract}

\section{Introduction}\label{sec:introduction}
Although the rank has long been known to be a metric implicitly and
explicitly (see, for example, \cite{Hua51}), the rank metric was
first considered for error control codes (ECCs) by Delsarte
\cite{delsarte_78}. The potential applications of rank metric codes
to wireless communications \cite{tarokh_98,lusina_it03}, public-key
cryptosystems \cite{gabidulin_lncs91}, and storage equipments
\cite{gabidulin_pit0285,roth_it91} have motivated a steady stream of
works \cite{gabidulin_pit0185, gabidulin_pit0285, roth_it91,
babu_95, chen_mn96, roth_it97, vasantha_gs99, kshevetskiy_isit05,
gabidulin_isit05, gadouleau_globecom06, gadouleau_itw06,
gadouleau_it06, loidreau_07}, described below, that focus on their
properties.

The majority of previous works focus on rank distance properties,
code construction, and efficient decoding of rank metric codes, and
the seminal works in \cite{delsarte_78, gabidulin_pit0185,
roth_it91} have made significant contribution to these topics.
Independently in \cite{delsarte_78, gabidulin_pit0185, roth_it91}, a
Singleton bound (up to some variations) on the minimum rank distance
of codes was established, and a class of codes achieving the bound
with equality was constructed. We refer to this class of codes as
Gabidulin codes henceforth. In \cite{delsarte_78,
gabidulin_pit0185}, analytical expressions to compute the weight
distribution of linear codes achieving the Singleton bound with
equality were also derived. In \cite{gabidulin_pit0285}, it was
shown that Gabidulin codes are optimal for correcting crisscross
errors (referred to as lattice-pattern errors in
\cite{gabidulin_pit0285}). In \cite{roth_it91}, it was shown that
Gabidulin codes are also optimal in the sense of a Singleton bound
in crisscross weight, a metric
considered in \cite{roth_it91,roth_it97,SE05} for crisscross errors.
Decoding algorithms that parallel the extended Euclidean algorithm
and the Peterson-Gornstein-Zierler algorithm were introduced for
Gabidulin codes in \cite{gabidulin_pit0185} and \cite{roth_it91},
respectively. In \cite{delsarte_78}, the counterpart of the
MacWilliams identity, which relates the rank distance enumerator of
a code to that of its dual code, was established using association
schemes. Following the works in \cite{delsarte_78,
gabidulin_pit0185, roth_it91}, the construction in
\cite{gabidulin_pit0185} was extended in \cite{kshevetskiy_isit05}
and the properties of subspace subcodes and subfield subcodes were
considered in \cite{gabidulin_isit05, gabidulin_it07}; the
counterparts of the Welch-Berlekamp algorithm and the
Berlekamp-Massey algorithm were considered in \cite{loidreau_05} and
\cite{richter_isit04} respectively for Gabidulin codes; the error
performance of Gabidulin codes was investigated in \cite{roth_it97,
gadouleau_itw06, gadouleau_it06}.

Some previous works focus on the packing and covering properties of
rank metric codes. Both packing and covering properties are
significant for ECCs, and packing and covering radii are basic
geometric parameters of a code, important in several respects
\cite{delsarte_73}. For instance, the covering radius can be viewed
as a measure of performance: if the code is used for error
correction, then the covering radius is the maximum weight of a
correctable error vector \cite{berger_book71}; if the code is used
for data compression, then the covering radius is a measure of the
maximum distortion \cite{berger_book71}. The Hamming packing and
covering radii of ECCs have been extensively studied (see, for
example,\cite{macwilliams_77,blahut_83,cohen_book97}), whereas the
rank packing and covering radii have received relatively little
attention. It was shown that nontrivial perfect rank metric codes do
not exist in \cite{babu_95, chen_mn96,gadouleau_globecom06}. In
\cite{vasantha_gs99}, a sphere covering bound for rank metric codes
was introduced. Generalizing the concept of rank covering radius,
the multi-covering radii of codes with the rank metric were defined
in \cite{vasantha_itw1002}.  Bounds on the volume of balls with rank
radii were also derived \cite{loidreau_07}.

In this paper, we investigate packing, covering, and rank distance
properties of rank metric codes. The main contributions of this
paper are:
\begin{itemize}


\item In Section~\ref{sec:technical_results}, we establish further properties of elementary linear subspaces
(ELS's) \cite{gadouleau_it06}, and investigate properties of balls
with rank radii. In particular, we derive both upper and lower
bounds on the volume of balls with given rank radii that are tighter
than their respective counterpart in \cite{loidreau_07}. These
technical results are used later in our investigation of the sphere
covering properties of rank metric codes.

\item In Section~\ref{sec:packing}, we study the packing
properties of rank metric codes, and also derive the asymptotic
maximum code rate for a code with given relative minimum rank
distance.

\item In Section~\ref{sec:covering}, we study the covering properties of rank metric codes, and derive both
upper and lower bounds on the minimal cardinality of a code with
given length and rank covering radius. Our new bounds are tighter
than the bounds introduced in \cite{vasantha_gs99}. Using the sphere
covering bound, we also establish additional sphere covering
properties for linear rank metric codes, and prove that some classes
of rank metric codes have maximal covering radius. Finally, we
establish the asymptotic minimum code rate for a code with given
relative covering radius.


\item In Section~\ref{sec:macwilliams}, we study the rank weight properties of linear codes. We show that,
similar to the MacWilliams identity for the Hamming metric, the rank
weight distribution of any linear code can be expressed as an
analytical expression of that of its dual code. It is also
remarkable that our MacWilliams identity for the rank metric has a
similar form to that for the Hamming metric. Despite the similarity,
our new identity is proved using a different approach based on
linear spaces. The intermediate results obtained using our approach
offer interesting insight. We also derive identities that relate
moments of the rank weight distribution of a
linear code to those of its dual code.

\end{itemize}


We provide the following remarks on our results:
\begin{enumerate}\item The concept of elementary
linear subspace was introduced in our previous work
\cite{gadouleau_it06}. It has similar properties to those of a set
of coordinates, and as such has served as a useful tool in our
derivation of properties of the rank metric (see
Section~\ref{sec:technical_results}), covering properties of rank
metric codes in general (see Section~\ref{sec:covering}), and
Gabidulin codes (see \cite{gadouleau_it06}). Although our results
may be derived without the concept of ELS, we have adopted it in
this paper since it enables readers to easily relate our approach
and results to their counterparts for Hamming metric codes.
\item Both the matrix form \cite{delsarte_78,roth_it91} and the vector form
\cite{gabidulin_pit0185} for rank metric codes have been considered
in the literature. Following \cite{gabidulin_pit0185}, in this paper
the vector form over $\mathrm{GF}(q^m)$ is used for rank metric
codes although their rank weight is defined by their corresponding
$m \times n$ code matrices over $\mathrm{GF}(q)$
\cite{gabidulin_pit0185}. The vector form is chosen in this paper
since our results and their derivations for rank metric codes can be
related to their counterparts for Hamming metric codes. \item In
\cite{delsarte_78}, the MacWilliams identity is given between the
rank distance enumerator sequences of two dual codes using the
generalized Krawtchouk polynomials. Based on a different proof, we
establish the same identity for linear rank metric codes, although
our identity is expressed using different parameters which are shown
to be the generalized Krawtchouk polynomials as well. We also
present this identity in weight enumerator polynomial form (cf.
Theorem~\ref{th:MacWilliams}). In their polynomial forms, the
MacWilliams identities for both the rank and the Hamming metrics are
similar to each other. Furthermore, the polynomial form allows us to
derive further identities (cf. Propositions~\ref{prop:bm_x} and
\ref{prop:bm_y}) between the rank weight distribution of linear dual
codes.
\end{enumerate}

The rest of the paper is organized as follows.
Section~\ref{sec:preliminaries} gives a brief review of necessary
background to keep this paper self-contained. In
Section~\ref{sec:technical_results}, we derive some further
properties of ELS's and balls of rank radii. In
Sections~\ref{sec:packing} and \ref{sec:covering}, we investigate
the packing and covering properties respectively of rank metric
codes. Section~\ref{sec:macwilliams} presents our results on the
rank weight distribution of rank metric codes.

\section{Preliminaries}\label{sec:preliminaries}
\subsection{Rank metric and elementary linear subspaces}\label{sec:rank_metric}
Consider an $n$-dimensional vector ${\bf x} = (x_0, x_1,\ldots,
x_{n-1}) \in \mathrm{GF}(q^m)^n$. The field $\mathrm{GF}(q^m)$ may
be viewed as an $m$-dimensional vector space over $\mathrm{GF}(q)$.
The rank weight of ${\bf x}$, denoted as $\rk({\bf x})$, is defined
to be the \emph{maximum} number of coordinates in ${\bf x}$ that are
linearly independent over $\mathrm{GF}(q)$ \cite{gabidulin_pit0185}.
Note that all ranks are with respect to $\mathrm{GF}(q)$ unless
otherwise specified in this paper. The coordinates of ${\bf x}$ thus
span a linear subspace of $\mathrm{GF}(q^m)$, denoted as
$\mathfrak{S}({\bf x})$, with dimension equal to $\rk({\bf x})$. For
all ${\bf x}, {\bf y}\in \mathrm{GF}(q^m)^n$, it is easily verified
that $\dr({\bf x},{\bf y})\df \rk({\bf x} - {\bf y})$ is a metric
over GF$(q^m)^n$, referred to as the \emph{rank metric} henceforth
\cite{gabidulin_pit0185}. The {\em minimum rank distance} of a code
$C$, denoted as $d_{\mbox{\tiny R}}(C)$, is simply the minimum rank
distance over all possible pairs of distinct codewords. When there
is no ambiguity about $C$, we denote the minimum rank distance as
$\dr$.

In \cite{gadouleau_it06}, we introduced the concept of elementary
linear subspace (\ELS{}). If there exists a basis set $B$ of vectors
in $\mathrm{GF}(q)^n$ for a linear subspace $\mathcal{V} \subseteq
\mathrm{GF}(q^m)^n$, we say $\mathcal{V}$ is an elementary linear
subspace and $B$ is an elementary basis of $\mathcal{V}$. We denote
the set of all \ELS{}'s of $\mathrm{GF}(q^m)^n$ with dimension $v$
as $E_v(q^m,n)$. An ELS has properties similar to those for a set of
coordinates \cite{gadouleau_it06}, and they are summarized as
follows. A vector has rank $\leq r$ if and only if it belongs to
some ELS with dimension $r$. For any $\mathcal{V} \in E_v(q^m,n)$,
there exists $\bar{\mathcal{V}} \in E_{n-v}(q^m,n)$ such that
$\mathcal{V} \oplus \bar{\mathcal{V}} = \mathrm{GF}(q^m)^n$, where
$\oplus$ denotes the direct sum of two subspaces. For any vector
${\bf x} \in \mathrm{GF}(q^m)^n$, we denote the projection of ${\bf
x}$ on $\mathcal{V}$ along $\bar{\mathcal{V}}$ as ${\bf
x}_\mathcal{V}$, and we remark that ${\bf x} = {\bf x}_\mathcal{V} +
{\bf x}_{\bar{\mathcal{V}}}$.

\subsection{The Singleton bounds}\label{sec:singleton}

It can be shown that $d_{\mbox{\tiny R}}\leq d_{\mbox{\tiny H}}$
\cite{gabidulin_pit0185}, where $d_{\mbox{\tiny H}}$ is the minimum
Hamming distance of the same code. Due to the Singleton bound for
block codes, the minimum rank distance of an $(n,k)$ block code over
$\mathrm{GF}(q^m)$ thus satisfies
\begin{equation}\label{eq:singleton1}
    d_{\mbox{\tiny R}}\leq n-k+1.
\end{equation}
An alternative bound on the minimum rank distance is also given in
\cite{loidreau_01}:
    \begin{equation}\label{eq:singleton2}
        d_{\mbox{\tiny R}} \leq \frac{m}{n}(n-k) + 1.
    \end{equation}
For $n \leq m$, the bound in (\ref{eq:singleton1}) is tighter than
that in (\ref{eq:singleton2}). When $n>m$ the bound in
(\ref{eq:singleton2}) is tighter.

When $n \leq m$, a class of codes satisfying~(\ref{eq:singleton1})
with equality was first proposed in \cite{gabidulin_pit0185} and
then generalized in \cite{kshevetskiy_isit05}. Let $\mathbf{g} =
(g_0, g_1, \ldots, g_{n-1})$ be linearly independent elements of
$\mathrm{GF}(q^m)$, then the code defined by the generator matrix
\begin{equation}
    \mathbf{G} = \left( \begin{array}{cccc}
    g_0 & g_1 & \ldots & g_{n-1}\\
    g_0^{[1]} & g_1^{[1]} & \ldots & g_{n-1}^{[1]}\\
    \vdots & \vdots & \ddots & \vdots\\
    g_0^{[k-1]} & g_1^{[k-1]} & \ldots & g_{n-1}^{[k-1]}
    \end{array}
    \right),\label{eq:gabidulin}
\end{equation}
where $[i] = q^{ai}$ with $a$ being an integer prime to $m$, is
called a generalized Gabidulin code generated by $\mathbf{g} = (g_0,
g_1, \ldots, g_{n-1})$; it has dimension $k$ and minimum rank
distance $d_{\mbox{\tiny R}} = n-k+1$ \cite{kshevetskiy_isit05}.

A class of codes satisfying~(\ref{eq:singleton2}) with equality was
proposed in \cite{loidreau_01}. It consists of cartesian products of
a generalized Gabidulin code with length $n = m$. Let $\mathcal{G}$
be an $(m, k, d_{\mbox{\tiny R}}=m-k+1)$ generalized Gabidulin code
over GF$(q^m)$, and let $\mathcal{G}^l \df \mathcal{G} \times \ldots
\times \mathcal{G}$ be the code obtained by $l$ cartesian products
of $\mathcal{G}$. Thus $\mathcal{G}^l$ is a code over
$\mathrm{GF}(q^m)$ with length $ml$, dimension $kl$, and minimum
rank distance $\dr = m-k+1$ \cite{loidreau_01}.


\subsection{Covering radius and excess}\label{sec:covering_radius}
The covering radius $\rho$ of a code $C$ with length $n$ over
$\mathrm{GF}(q^m)$ is defined to be the smallest integer $\rho$ such
that all vectors in the space $\mathrm{GF}(q^m)^n$ are within
distance $\rho$ of some codeword of $C$ \cite{cohen_book97}. It is
the maximal distance from any vector in $\mathrm{GF}(q^m)^n$  to the
code $C$. That is, $\rho = \max_{{\bf x} \in \mathrm{GF}(q^m)^n} \{
d({\bf x},C)\}$. Also, if $C \subset C'$, then the covering radius
of $C$ is no more than the minimum distance of $C'$. Finally, a code
$C$ with length $n$ and minimum distance $d$ is called a maximal
code if there does not exist any code $C'$ with same length and
minimum rank distance such that $C \subset C'$. A maximal code has
covering radius $\rho \leq d-1$.

Van Wee \cite{vanwee_88,vanwee_91} derived several bounds on codes
with Hamming covering radii based on the excess of a code, which is
determined by the number of codewords covering the same vectors.
Below are some key definitions and results in
\cite{vanwee_88,vanwee_91}. For all $V \subseteq \mathrm{GF}(q^m)^n$
and a code $C$ with covering radius $\rho$, the excess on $V$ by $C$
is defined to be
\begin{equation}
    E_C(V) \df \sum_{{\bf c} \in C}|B^{\mbox{\tiny H}}_\rho({\bf c}) \cap V| -
    |V|,
\end{equation}
where $B^{\mbox{\tiny H}}_\rho({\bf c})$ denotes a ball centered at
${\bf c}$ with Hamming radius $\rho$. The excess on
$\mathrm{GF}(q^m)^n$ by $C$ is given by $E_C(\mathrm{GF}(q^m)^n) =
|C| \cdot V^{\mbox{\tiny H}}_{\rho}(q^m,n) - q^{mn}$, where
$V^{\mbox{\tiny H}}_{\rho}(q^m,n)$ denotes the volume of a ball with
Hamming radius $\rho$. Also, if $\{W_i\}$ is a family of disjoint
subsets of $\mathrm{GF}(q^m)^n$, then $E_C \left( \bigcup_i W_i
\right) = \sum_i E_C(W_i)$. Suppose $Z \df \{{\bf z} \in
\mathrm{GF}(q^m)^n | E_C(\left\{{\bf z}\right\}) > 0 \}$
\cite{vanwee_88}, i.e., $Z$ is the set of vectors covered by at
least two codewords in $C$. Note that ${\bf z} \in Z$ if and only if
$|B^{\mbox{\tiny H}}_{\rho}({\bf z}) \cap C| \geq 2$. It can be
shown that $|Z| \leq E_C(Z) = E_C(\mathrm{GF}(q^m)^n) = |C|\cdot
V^{\mbox{\tiny H}}_{\rho}(q^m,n) - q^{mn}$.

Although the above definitions and properties were developed for the
Hamming metric, they are in fact independent of the underlying
metric and thus are applicable to the rank metric as well.

\subsection{Notations}\label{sec:notations}
In order to simplify notations, we shall occasionally denote the
vector space $\mathrm{GF}(q^m)^n$ as $F$. We denote the number of
vectors of rank $u$ ($0 \leq u \leq \min\{m,n\}$) in
$\mathrm{GF}(q^m)^n$ as $N_u(q^m,n)$. It can be shown that
$N_u(q^m,n) = {n \brack u} \alpha(m,u)$ \cite{gabidulin_pit0185},
where $\alpha(m,0) \df 1$ and $\alpha(m,u) \df
\prod_{i=0}^{u-1}(q^m-q^i)$ for $u \geq 1$. The ${n \brack u}$ term
is often referred to as a Gaussian polynomial~\cite{andrews},
defined as ${n \brack u} \df \alpha(n,u)/\alpha(u,u)$. Note that ${n
\brack u}$ is the number of $u$-dimensional linear subspaces of
$\mathrm{GF}(q)^n$. We refer to all vectors in $\mathrm{GF}(q^m)^n$
within rank distance $r$ of ${\bf x} \in \mathrm{GF}(q^m)^n$ as a
ball of rank radius $r$ centered at ${\bf x}$, and denote it as
$B_r({\bf x})$. Its volume, which does not depend on ${\bf x}$, is
denoted as $V_r(q^m,n) = \sum_{u=0}^r N_u(q^m,n)$. We also define
$\beta(m,0) \df 1$ and $\beta(m,u) \df \prod_{i=0}^{u-1} {m-i \brack
1}$ for $u \geq 1$, which are used in Section~\ref{sec:macwilliams}.
These terms are closely related to Gaussian polynomials: $\beta(m,u)
= {m \brack u} \beta(u,u)$ and $\beta(m+u,m+u) = {m+u \brack u}
\beta(m,m) \beta(u,u)$.

\section{Technical results}\label{sec:technical_results}
\subsection{Further properties of \ELS{}'s}\label{sec:lemmas_ELS}

%


%


\begin{lemma}\label{lemma:unique_ELS}
Any vector ${\bf x} \in \mathrm{GF}(q^m)^n$ with rank $r$ belongs to
a unique \ELS{} $\mathcal{V} \in E_r(q^m,n)$.
\end{lemma}

\begin{proof}
The existence of $\mathcal{V} \in E_r(q^m,n)$ has been proved in
\cite{gadouleau_it06}. Thus we only prove the uniqueness of
$\mathcal{V}$, with elementary basis $\{ {\bf v}_i \}_{i=0}^{r-1}$.
Suppose ${\bf x}$ also belongs to $\mathcal{W}$, where $\mathcal{W}
\in E_r(q^m,n)$ has an elementary basis $\{ {\bf w}_j
\}_{j=0}^{r-1}$. Therefore, ${\bf x} = \sum_{i=0}^{r-1} a_i {\bf
v}_i = \sum_{j=0}^{r-1} b_j {\bf w}_j$, where $a_i,b_j \in
\mathrm{GF}(q^m)$ for $0 \leq i,j \leq r-1$. By definition, we have
$\mathfrak{S}({\bf x}) = \mathfrak{S}(a_0,\ldots,a_{r-1}) =
\mathfrak{S}(b_0,\ldots,b_{r-1})$, therefore $b_j$'s can be
expressed as linear combinations of $a_i$'s, i.e., $b_j =
\sum_{i=0}^{r-1} c_{j,i} a_i$ where $c_{j,i} \in \mathrm{GF}(q)$.
Hence
\begin{equation}\label{eq:x=sum_u}
    {\bf x}  = \sum_{j=0}^{r-1} b_j {\bf w}_j
    = \sum_{j=0}^{r-1} \sum_{i=0}^{r-1} c_{j,i} a_i {\bf w}_j
    = \sum_{i=0}^{r-1} a_i {\bf u}_i,
\end{equation}
where ${\bf u}_i = \sum_{j=0}^{r-1} c_{j,i} {\bf w}_j \in
\mathrm{GF}(q)^n$. Now consider ${\bf X}$, the matrix obtained by
expanding the coordinates of ${\bf x}$ with respect to the basis
$\{a_i\}_{i=0}^{m-1}$. For $0 \leq i \leq r-1$, the $i$-th row of
${\bf X}$ is given by the vector ${\bf v}_i$ by definition and by
the vector ${\bf u}_i$ from Eq.~(\ref{eq:x=sum_u}). Therefore ${\bf
v}_i = {\bf u}_i \in \mathcal{W}$, and hence $\mathcal{V} \subseteq
\mathcal{W}$. However, $\dim(\mathcal{V}) = \dim(\mathcal{W})$, and
thus $\mathcal{V} = \mathcal{W}$.
\end{proof}

Lemma~\ref{lemma:unique_ELS} shows that an ELS is analogous to a
subset of coordinates since a vector ${\bf x}$ with Hamming weight
$r$ belongs to a unique subset of $r$ coordinates, often referred to
as the support of ${\bf x}$.

In \cite{gadouleau_it06}, it was shown that an \ELS{} always has a
complementary elementary linear subspace. The following lemma
enumerates such complementary ELS's.
\begin{lemma}\label{lemma:num_B}
Suppose $\mathcal{V} \in E_v(q^m,n)$ and $\mathcal{A} \subseteq
\mathcal{V}$ is an \ELS{} with dimension $a$, then there are
$q^{a(v-a)}$ \ELS{}'s $\mathcal{B}$ such that $\mathcal{A} \oplus
\mathcal{B} = \mathcal{V}$. Furthermore, there are $q^{a(v-a)} {v
\brack a}$ such ordered pairs $(\mathcal{A},\mathcal{B})$.
\end{lemma}

\begin{proof}
First, remark that $\mathrm{dim}(\mathcal{B}) = v-a$. The total
number of sets of $v-a$ linearly independent vectors over GF$(q)$ in
$\mathcal{V} \backslash \mathcal{A}$ is given by $N = (q^v -
q^a)(q^v-q^{a+1}) \cdots (q^v-q^{v-1}) = q^{a(v-a)}\alpha(v-a,v-a)$.
Note that each set of linearly independent vectors over GF$(q)$
constitutes an elementary basis set. Thus, the number of possible
$\mathcal{B}$ is given by $N$ divided by $\alpha(v-a,v-a)$, the
number of elementary basis sets for each $\mathcal{B}$. Therefore,
once $\mathcal{A}$ is fixed, there are $q^{a(v-a)}$ choices for
$\mathcal{B}$. Since the number of $a$-dimensional subspaces
$\mathcal{A}$ in $\mathcal{V}$ is ${v \brack a}$, the total number
of ordered pairs $(\mathcal{A},\mathcal{B})$ is hence $q^{a(v-a)} {v
\brack a}$.
\end{proof}

Puncturing a vector with full Hamming weight results in another
vector with full Hamming weight. Lemma~\ref{lemma:restriction_u}
below shows that the situation for vectors with full rank is
similar.

\begin{lemma}\label{lemma:restriction_u}
Suppose $\mathcal{V} \in E_v(q^m,n)$ and ${\bf u} \in \mathcal{V}$
has rank $v$, then $\rk({\bf u}_\mathcal{A}) = a$ and $\rk({\bf
u}_\mathcal{B}) = v-a$ for any $\mathcal{A} \in E_a(q^m,n)$ and
$\mathcal{B} \in E_{v-a}(q^m,n)$ such that $\mathcal{A} \oplus
\mathcal{B} = \mathcal{V}$.
\end{lemma}

\begin{proof}
First, ${\bf u}_\mathcal{A} \in \mathcal{A}$ and hence $\rk({\bf
u}_\mathcal{A}) \leq a$ by \cite[Proposition~2]{gadouleau_it06};
similarly, $\rk({\bf u}_\mathcal{B}) \leq v-a$. Now suppose
$\rk({\bf u}_\mathcal{A}) < a$ or $\rk({\bf u}_\mathcal{B}) < v-a$,
then $v = \rk({\bf u}) \leq \rk({\bf u}_\mathcal{A}) + \rk({\bf
u}_\mathcal{B}) < a+v-a = v$.
\end{proof}

It was shown in \cite{gadouleau_it06} that the projection ${\bf
u}_\mathcal{A}$ of a vector ${\bf u}$ on an \ELS{} $\mathcal{A}$
depends on both $\mathcal{A}$ and its complement $\mathcal{B}$. The
following lemma further clarifies the relationship: changing
$\mathcal{B}$ always modifies ${\bf u}_\mathcal{A}$, provided that
${\bf u}$ has full rank.

\begin{lemma}\label{lemma:restriction}
Suppose $\mathcal{V} \in E_v(q^m,n)$ and ${\bf u} \in \mathcal{V}$
has rank $v$.  For any $\mathcal{A} \in E_a(q^m,n)$ and $\mathcal{B}
\in E_{v-a}(q^m,n)$ such that $\mathcal{A} \oplus \mathcal{B} =
\mathcal{V}$, define the functions $f_{\bf
u}(\mathcal{A},\mathcal{B}) = {\bf u}_\mathcal{A}$ and $g_{\bf
u}(\mathcal{A},\mathcal{B}) ={\bf u}_\mathcal{B}$. Then both $f_{\bf
u}$ and $g_{\bf u}$ are injective.
\end{lemma}

\begin{proof}
Consider another pair $(\mathcal{A}',\mathcal{B}')$ with dimensions
$a$ and $v-a$ respectively. Suppose $\mathcal{A}' \neq \mathcal{A}$,
then ${\bf u}_{\mathcal{A}'} \neq {\bf u}_\mathcal{A}$. Otherwise
${\bf u}_\mathcal{A}$ belongs to two distinct \ELS{}'s with
dimension $a$, which contradicts Lemma~\ref{lemma:unique_ELS}. Hence
${\bf u}_{\mathcal{A}'} \neq {\bf u}_\mathcal{A}$ and ${\bf
u}_{\mathcal{B}'} = {\bf u} - {\bf u}_{\mathcal{A}'} \neq {\bf u} -
{\bf u}_{\mathcal{A}} = {\bf u}_\mathcal{B}$. The argument is
similar if $\mathcal{B}' \neq \mathcal{B}$.
\end{proof}

\subsection{Properties of balls with rank radii}\label{sec:balls}

\begin{lemma}\label{lemma:lower_bound_Vt}
For $0 \leq r \leq \min\{n,m\}$,
\begin{equation}
    q^{r(m+n-r)} \leq V_r(q^m,n) < q^{r(m+n-r) + \sigma(q)},
    \label{eq:lower_bound_Vt}
\end{equation}
where $\sigma(q) \df \frac{1}{\ln(q)} \sum_{k=1}^\infty
\frac{1}{k(q^k-1)}$ is a decreasing function of $q$ satisfying
$\sigma(q) <2$ for $q \geq 2$ \cite{gadouleau_it06}.
\end{lemma}

\begin{proof}
The upper bound in (\ref{eq:lower_bound_Vt}) was derived in
\cite[Lemma 13]{gadouleau_it06}, and it suffices to prove the lower
bound. Without loss of generality, we assume that the center of the
ball is ${\bf 0}$. For any ${\bf x} \in \mathrm{GF}(q^m)^r$, we
associate one subspace $\mathfrak{T}$ of $\mathrm{GF}(q^m)$ such
that $\mathrm{dim}(\mathfrak{T}) = r$ and $\mathfrak{S}({\bf x})
\subseteq \mathfrak{T}$. We consider the vectors ${\bf y} \in
\mathrm{GF}(q^m)^{n-r}$ such that $\mathfrak{S}({\bf y}) \subseteq
\mathfrak{T}$. There are $q^{mr}$ choices for ${\bf x}$ and, for a
given ${\bf x}$, $q^{r(n-r)}$ choices for ${\bf y}$. Thus the total
number of vectors ${\bf z} = ({\bf x},{\bf y}) \in
\mathrm{GF}(q^m)^n$ is $q^{r(m+n-r)}$. Since $\mathfrak{S}({\bf z})
\subseteq \mathfrak{T}$, we have $\rk({\bf z}) \leq r$. Thus,
$V_r(q^m,n) \geq q^{r(m+n-r)}$.
\end{proof}

We remark that both bounds in (\ref{eq:lower_bound_Vt}) are tighter
than their respective counterparts in
\cite[Proposition~1]{loidreau_07}. More importantly, the two bounds
in (\ref{eq:lower_bound_Vt}) differ only by a factor of
$q^{\sigma(q)}$, and thus they not only provide a good approximation
of $V_r(q^m,n)$, but also accurately describe the asymptotic
behavior of $V_r(q^m,n)$.

The diameter of a set is defined to be the maximum distance between
any pair of elements in the set \cite[p. 172]{macwilliams_77}. For a
binary vector space $\mathrm{GF}(2)^n$ and a given diameter $2r<n$,
Kleitman \cite{kleitman_66} proved that balls with Hamming radius
$r$ maximize the cardinality of a set with a given diameter.
However, when the underlying field for the vector space is not
$\mathrm{GF}(2)$, the result is not necessarily valid \cite[p.
40]{cohen_book97}. We show below that balls with rank radii do not
necessarily maximize the cardinality of a set with a given diameter.

\begin{proposition}\label{prop:counter_Kleitman}
For $3 \leq n \leq m$ and $2 \leq 2r < n$, there exists $S \subset
\mathrm{GF}(q^m)^n$ with diameter $2r$ such that $|S| > V_r(q^m,n)$.
\end{proposition}

\begin{proof}
The set $S \df \{ (x_0,\ldots,x_{n-1}) \in \mathrm{GF}(q^m)^n |
x_{2r} = \cdots = x_{n-1} = 0\}$ has diameter $2r$ and cardinality
$q^{2mr}$. For $r=1$, we have $V_1(q^m,n) = 1 +
\frac{(q^n-1)(q^m-1)}{(q-1)} < q^{2m}$. For $r \geq 2$, we have
$V_r(q^m,n) < q^{r(n+m)-(r^2-\sigma(q))}$ by
Lemma~\ref{lemma:lower_bound_Vt}. Since $r^2 > 2 > \sigma(q)$, we
obtain $V_r(q^m,n) < q^{r(n+m)} \leq |S|$.
\end{proof}

The intersection of balls with Hamming radii has been studied in
\cite[Chapter 2]{cohen_book97}, and below we investigate the
intersection of balls with rank radii.

\begin{lemma}\label{lemma:inter_2_balls}
If $0 \leq r,s \leq n$ and ${\bf c}_1, {\bf c}_2 \in
\mathrm{GF}(q^m)^n$, then $|B_r({\bf c}_1) \cap B_s({\bf c}_2)|$
depends on ${\bf c}_1$ and ${\bf c}_2$ only through $\dr({\bf
c}_1,{\bf c}_2)$.
\end{lemma}

\begin{proof}
First, without loss of generality, we assume ${\bf c}_1 = {\bf 0}$,
and we denote $\rk({\bf c}_2) = e$. We can express ${\bf c}_2$ as
${\bf c}_2 = {\bf u} {\bf B}$, where ${\bf u} =
(u_0,\ldots,u_{e-1},0,\ldots,0) \in \mathrm{GF}(q^m)^n$ has rank $e$
and ${\bf B} \in \mathrm{GF}(q)^{n \times n}$ has full rank. For any
${\bf x} \in B_r({\bf 0}) \cap B_s({\bf u})$ we have $\rk({\bf
x}{\bf B}) = \rk({\bf x}) \leq r$ and $\rk({\bf x}{\bf B} - {\bf
c}_2) = \rk({\bf x} - {\bf u}) \leq s$. Thus there is a bijection
between $B_r({\bf 0}) \cap B_s({\bf u}{\bf B})$ and $B_r({\bf 0})
\cap B_s({\bf u})$. Hence $|B_r({\bf 0}) \cap B_s({\bf u} {\bf B})|
= |B_r({\bf 0}) \cap B_s({\bf u})|$, that is, $|B_r({\bf 0}) \cap
B_s({\bf u} {\bf B})|$ does not depend on ${\bf B}$.

Since $|B_r({\bf 0}) \cap B_s({\bf u} {\bf B})|$ is independent of
${\bf B}$, we assume ${\bf B} = {\bf I}_{n \times n}$ without loss
of generality henceforth. The nonzero coordinates of ${\bf u}$ all
belong to a basis set $\{u_i\}_{i=0}^{m-1}$ of $\mathrm{GF}(q^m)$.
Let ${\bf x} = (x_0,\ldots,x_{n-1}) \in B_r({\bf 0}) \cap B_s({\bf
u})$, then we can express $x_j$ as $x_j = \sum_{i=0}^{m-1}
a_{i,j}u_i$ with $a_{i,j} \in \mathrm{GF}(q)$ for $0 \leq j \leq
n-1$. Suppose ${\bf v} = (v_0,\ldots,v_{e-1},0,\ldots,0) \in
\mathrm{GF}(q^m)^n$ also has rank $e$, then the nonzero coordinates
of ${\bf v}$ all belong to a basis set $\{v_i\}_{i=0}^{m-1}$ of
$\mathrm{GF}(q^m)$. We define $\bar{\bf x} =
(\bar{x}_0,\ldots,\bar{x}_{n-1}) \in \mathrm{GF}(q^m)^n$ such that
$\bar{x}_j = \sum_{j=0}^{m-1} a_{i,j}v_i$ for $0 \leq j \leq n-1$.
We remark that $\rk(\bar{\bf x}) = \rk({\bf x}) \leq r$ and
$\rk(\bar{\bf x} - {\bf v}) = \rk({\bf x} - {\bf u}) \leq s$. Thus
there is a bijection between $B_r({\bf 0}) \cap B_s({\bf v})$ and
$B_r({\bf 0}) \cap B_s({\bf u})$. Hence $|B_r({\bf 0}) \cap B_s({\bf
u})|$ depends on the vector ${\bf u}$ only through its rank $e$.
\end{proof}

\begin{proposition}\label{prop:inter_2_balls}
If $0 \leq r,s \leq n$, ${\bf c}_1, {\bf c}_2, {\bf c}'_1, {\bf
c}'_2 \in \mathrm{GF}(q^m)^n$ and $\dr({\bf c}_1,{\bf c}_2)
> \dr({\bf c}'_1,{\bf c}'_2)$, then
\begin{equation}\label{eq:inter_2_balls}
    |B_r({\bf c}_1) \cap B_s({\bf c}_2)| \leq
    |B_r({\bf c}'_1) \cap B_s({\bf c}'_2)|.
\end{equation}
\end{proposition}

\begin{proof}
It suffices to prove~(\ref{eq:inter_2_balls}) when $\dr({\bf
c}_1,{\bf c}_2) = \dr({\bf c}'_1,{\bf c}'_2) + 1 = e + 1$. By
Lemma~\ref{lemma:inter_2_balls}, we can assume without loss of
generality that ${\bf c}_1 = {\bf c}'_1 = {\bf 0}$, ${\bf c}'_2 =
(0, c_1,\ldots,c_e,0,\ldots,0)$ and ${\bf c}_2 = (c_0,
c_1,\ldots,c_e,0,\ldots,0)$, where $c_0,\ldots,c_e \in
\mathrm{GF}(q^m)$ are linearly independent.

We will show that an injective mapping $\phi$ from $B_r({\bf c}_1)
\cap B_s({\bf c}_2)$ to $B_r({\bf c}'_1) \cap B_s({\bf c}'_2)$ can
be constructed. We consider vectors ${\bf z} =
(z_0,z_1,\ldots,z_{n-1}) \in B_r({\bf c}_1) \cap B_s({\bf c}_2)$. We
thus have $\rk({\bf z}) \leq r$ and $\rk({\bf u}) \leq s$, where
${\bf u} = (u_0,u_1,\ldots,u_{n-1}) = {\bf z} - {\bf c}_2 =
(z_0-c_0,z_1-c_1,\ldots,z_{n-1})$. We also define $\bar{\bf z} =
(z_1,\ldots,z_{n-1})$ and $\bar{\bf u} = (u_1,\ldots,u_{n-1})$. We
consider three cases for the mapping $\phi$, depending on $\bar{\bf
z}$ and $\bar{\bf u}$.
\begin{itemize}
\item Case I: $\rk(\bar{\bf u}) \leq s-1$. In this case, $\phi({\bf z}) \df {\bf z}$.
We remark that $\rk({\bf z} - {\bf c}'_2) \leq \rk(\bar{\bf u}) + 1
\leq s$ and hence $\phi({\bf z}) \in B_r({\bf c}'_1) \cap B_s({\bf
c}'_2)$.

\item Case II: $\rk(\bar{\bf u}) = s$ and $\rk(\bar{\bf z}) \leq r-1$.
In this case, $\phi({\bf z}) \df (z_0-c_0,z_1,\ldots,z_{n-1})$. We
have $\rk(\phi({\bf z})) \leq \rk(\bar{\bf z}) +1 \leq r$ and
$\rk\left(\phi({\bf z}) - {\bf c}'_2\right) = \rk\left({\bf z}- {\bf
c}_2\right)\leq s$, and hence $\phi({\bf z}) \in B_r({\bf c}'_1)
\cap B_s({\bf c}'_2)$.

\item Case III: $\rk(\bar{\bf u}) = s$ and $\rk(\bar{\bf z}) = r$.
Since $\rk({\bf u}) = s$, we have $z_0 - c_0 \in
\mathfrak{S}(\bar{\bf u})$. Similarly, since $\rk({\bf z}) = r$, we
have $z_0 \in \mathfrak{S}(\bar{\bf z})$. Denote
$\dim(\mathfrak{S}(\bar{\bf u},\bar{\bf z}))$ as $d$ ($d\geq s$).
For $d
> s$, let $\alpha_0,\ldots,\alpha_{d-1}$ be a basis of
$\mathfrak{S}(\bar{\bf u},\bar{\bf z})$ such that
$\alpha_0,\ldots,\alpha_{s-1} \in \mathfrak{S}(\bar{\bf u})$ and
$\alpha_s,\ldots,\alpha_{d-1} \in \mathfrak{S}(\bar{\bf z})$. Note
that $c_0 \in \mathfrak{S}(\bar{\bf u},\bar{\bf z})$, and may
therefore be uniquely expressed as $c_0 = c_u + c_z$, where $c_u \in
\mathfrak{S}(\alpha_0,\ldots,\alpha_{s-1})\subseteq
\mathfrak{S}(\bar{\bf u})$ and $c_z \in
\mathfrak{S}(\alpha_s,\ldots,\alpha_{d-1})\subseteq
\mathfrak{S}(\bar{\bf z})$. If $d=s$, then $c_z=0\in
\mathfrak{S}(\bar{\bf z})$. In this case, $\phi({\bf z}) \df (z_0 -
c_z,z_1,\ldots,z_{n-1})$. Remark that $z_0-c_z \in
\mathfrak{S}(\bar{\bf z})$ and hence $\rk(\phi({\bf z})) = r$. Also,
$z_0-c_z = z_0 - c_0 + c_u \in \mathfrak{S}(\bar{\bf u})$ and hence
$\rk(\phi({\bf z}) - {\bf c}'_2) = s$. Therefore $\phi({\bf z}) \in
B_r({\bf c}'_1) \cap B_s({\bf c}'_2)$.
\end{itemize}

We now verify that the mapping $\phi$ is injective. Suppose there
exists ${\bf z}'$ such that $\phi({\bf z}') = \phi({\bf z})$. Since
$\phi({\bf z})$ only modifies the first coordinate of ${\bf z}$, the
last $n-1$ coordinates of ${\bf z}$ and ${\bf z}'$ are equal and so
are the last $n-1$ coordinates of ${\bf z} - {\bf c}_2$ and ${\bf
z}' - {\bf c}_2$. Hence ${\bf z}$ and ${\bf z}'$ belong to the same
case. It can be easily verified that for each case above, $\phi$ is
injective. Hence $\phi({\bf z}')= \phi({\bf z})$ implies that ${\bf
z}' = {\bf z}$. Therefore $\phi$ is injective, and $|B_r({\bf c}_1)
\cap B_s({\bf c}_2)| \leq |B_r({\bf c}'_1) \cap B_s({\bf c}'_2)|$.
\end{proof}

\begin{corollary}\label{cor:union_2_balls}
If $0 \leq r,s \leq n$, ${\bf c}_1, {\bf c}_2, {\bf c}'_1, {\bf
c}'_2 \in \mathrm{GF}(q^m)^n$ and $\dr({\bf c}_1,{\bf c}_2) \geq
\dr({\bf c}'_1,{\bf c}'_2)$, then
\begin{equation}
    |B_r({\bf c}_1) \cup B_s({\bf c}_2)| \geq
    |B_r({\bf c}'_1) \cup B_s({\bf c}'_2)|.
\end{equation}
\end{corollary}

\begin{proof}
The result follows from $|B_r({\bf c}_1) \cup B_s({\bf c}_2)| =
V_r(q^m,n) + V_s(q^m,n) - |B_r({\bf c}_1) \cap B_s({\bf c}_2)|$.
\end{proof}

We now quantify the volume of the intersection of two balls with
rank radii for some special cases, which will be used in
Section~\ref{sec:lower_bounds}.

\begin{proposition}\label{prop:B_r}
If ${\bf c}_1, {\bf c}_2 \in \mathrm{GF}(q^m)^n$ and $\dr({\bf c}_1,
{\bf c}_2) = r$, then $|B_r({\bf c}_1) \cap B_1({\bf c}_2)| = 1 +
(q^m-q^r){r \brack 1} + (q^r-1){n \brack 1}$.
\end{proposition}

\begin{proof}
 The claim holds for
$r=m$ trivially, and we assume $r<m$ henceforth. By
Lemma~\ref{lemma:inter_2_balls}, assume ${\bf c}_2 = {\bf 0}$ and
hence $\rk({\bf c}_1) = r$ without loss of generality. By
Lemma~\ref{lemma:unique_ELS}, the vector ${\bf c}_1$ belongs to a
unique \ELS{} $\mathcal{V} \in E_r(q^m,n)$. First of all, it is easy
to check that ${\bf y} = {\bf 0} \in B_r({\bf c}_1) \cap B_1({\bf
0})$. We consider a nonzero vector ${\bf y} \in B_1({\bf 0})$ with
rank $1$. Firstly, if ${\bf y} \in \mathcal{V}$, then ${\bf c}_1 -
{\bf y} \in \mathcal{V}$. We hence have $\rk({\bf c}_1 - {\bf y})
\leq r$ and ${\bf y} \in B_r({\bf c}_1)$. Note that there are
$(q^m-1){r \brack 1}$ such vectors. Secondly, if ${\bf y} \notin
\mathcal{V}$ and $\mathfrak{S}({\bf y}) \subseteq \mathfrak{S}({\bf
c}_1)$, then $\mathfrak{S}({\bf c}_1 - {\bf y}) \subseteq
\mathfrak{S}({\bf c}_1)$. We hence have $\rk({\bf c}_1 - {\bf y})
\leq r$ and ${\bf y} \in B_r({\bf c}_1)$. Note that there are
$(q^r-1)({n \brack 1} - {r \brack 1})$ such vectors. Finally,
suppose ${\bf y} \notin \mathcal{V}$ and $\mathfrak{S}({\bf y})
\nsubseteq \mathfrak{S}({\bf c}_1)$. Denote the linearly independent
coordinates of ${\bf c}_1$ as $\alpha_0,\ldots,\alpha_{r-1}$ and a
nonzero coordinate of ${\bf y}$ as $\alpha_r \notin
\mathfrak{S}({\bf c}_1)$, where $\{ \alpha_i \}_{i=0}^{m-1}$ is a
basis set of $\mathrm{GF}(q^m)$. Then the matrix ${\bf C}_1 - {\bf
Y}$ obtained by expanding the coordinates of ${\bf c}_1 - {\bf y}$
according to the basis $\{ \alpha_i\}$ has row rank $r+1$. Therefore
$\rk({\bf c}_1 - {\bf y}) = r+1$, and ${\bf y} \notin B_r({\bf
c}_1)$.
\end{proof}

\begin{proposition}\label{prop:v-a_problem}
If ${\bf c}_1, {\bf c}_2 \in \mathrm{GF}(q^m)^n$ and $\dr({\bf c}_1,
{\bf c}_2) = r$, then $|B_s({\bf c}_1) \cap B_{r-s}({\bf c}_2)| =
q^{s(r-s)} {r \brack s}$ for $0 \leq s \leq r$.
\end{proposition}


\begin{proof}
By Lemma~\ref{lemma:inter_2_balls}, we can assume that ${\bf c}_1 =
{\bf 0}$, and hence $\rk({\bf c}_2) = r$. By
Lemma~\ref{lemma:unique_ELS}, ${\bf c}_2$ belongs to a unique \ELS{}
$\mathcal{V} \in E_r(q^m,n)$. We first prove that all vectors ${\bf
y} \in B_s({\bf 0}) \cap B_{r-s}({\bf c}_2)$ are in $\mathcal{V}$.
Let ${\bf y} = {\bf y}_\mathcal{V} + {\bf y}_\mathcal{W}$, where
$\mathcal{W} \in E_{n-r}(q^m,n)$ such that $\mathcal{V} \oplus
\mathcal{W} = \mathrm{GF}(q^m)^n$. We have ${\bf y}_\mathcal{V} +
({\bf c}_2-{\bf y})_\mathcal{V} = {\bf c}_2$, with $\rk({\bf
y}_\mathcal{V}) \leq \rk({\bf y}) \leq s$ and $\rk(({\bf c}_2 - {\bf
y})_\mathcal{V}) \leq \rk({\bf c}_2 - {\bf y}) \leq r-s$. Therefore,
$\rk({\bf y}_\mathcal{V}) = \rk({\bf y}) = s$, $\rk(({\bf c}_2 -
{\bf y})_\mathcal{V}) = \rk({\bf c}_2 - {\bf y})= r-s$, and
$\mathfrak{S}({\bf y}_\mathcal{V}) \cap \mathfrak{S}(({\bf c}_2 -
{\bf y})_\mathcal{V}) = \{0 \}$. Since $\rk({\bf y}_\mathcal{V}) =
\rk({\bf y})$, we have $\mathfrak{S}({\bf y}_\mathcal{W}) \subseteq
\mathfrak{S}({\bf y}_\mathcal{V})$; and similarly
$\mathfrak{S}(({\bf c}_2 - {\bf y})_\mathcal{W}) \subseteq
\mathfrak{S}(({\bf c}_2 -{\bf y})_\mathcal{V})$. Altogether, we
obtain $\mathfrak{S}({\bf y}_\mathcal{W}) \cap \mathfrak{S}(({\bf
c}_2 - {\bf y})_\mathcal{W}) = \{0\}$. However, ${\bf y}_\mathcal{W}
+ ({\bf c}_2 - {\bf y})_\mathcal{W} = {\bf 0}$, and hence ${\bf
y}_\mathcal{W} = ({\bf c}_2 - {\bf y})_\mathcal{W} = {\bf 0}$.
Therefore, ${\bf y} \in \mathcal{V}$.

We now prove that ${\bf y}$ is necessarily the projection of ${\bf
c}_2$ onto some \ELS{} $\mathcal{A}$ of $\mathcal{V}$. If ${\bf y}
\in \mathcal{V}$ satisfies $\rk({\bf y}) = s$ and $\rk({\bf
c}_2-{\bf y}) = r-s$, then ${\bf y}$ belongs to some \ELS{}
$\mathcal{A}$ and ${\bf c}_2 - {\bf y} \in \mathcal{B}$ such that
$\mathcal{A} \oplus \mathcal{B} = \mathcal{V}$. We hence have ${\bf
y} = {\bf c}_{2,\mathcal{A}}$ and ${\bf c}_2 - {\bf y} = {\bf
c}_{2,\mathcal{B}}$.

On the other hand, for any $\mathcal{A} \in E_s(q^m,n)$ and
$\mathcal{B} \in E_{r-s}(q^m,n)$ such that $\mathcal{A} \oplus
\mathcal{B} = \mathcal{V}$, ${\bf c}_{2,\mathcal{A}}$ is a vector of
rank $s$ with distance $r-s$ from ${\bf c}_2$ by
Lemma~\ref{lemma:restriction_u}. By Lemma~\ref{lemma:restriction},
all the ${\bf c}_{2,\mathcal{A}}$ vectors are distinct. There are
thus as many vectors ${\bf y}$ as ordered pairs
$(\mathcal{A},\mathcal{B})$. By Lemma~\ref{lemma:num_B}, there are
$q^{s(r-s)}{r \brack s}$ such pairs, and hence $q^{s(r-s)}{r \brack
s}$ vectors ${\bf y}$.
\end{proof}

\begin{figure}[htp]
\begin{center}
\includegraphics[scale=1.0]{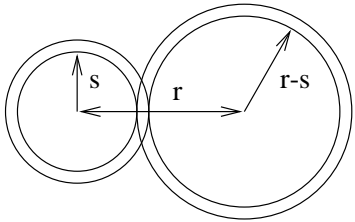}
\end{center}
\caption{Illustration of
Proposition~\ref{prop:v-a_problem}.}\label{fig:overlap}
\end{figure}
As shown in Figure~\ref{fig:overlap}, only the outmost layers of two
balls of radii $s$ and $r-s$ intersect when the distance between the
two centers is $r$. Proposition~\ref{prop:v-a_problem} quantifies
the volume of the intersection in Figure~\ref{fig:overlap}.

The problem of the intersection of three balls with rank radii is
more complicated since the volume of the intersection of three balls
with rank radii is not completely determined by the pairwise
distances between the centers. We give a simple example to
illustrate this point: consider $\mathrm{GF}(2^2)^3$ and the vectors
${\bf c}_1 = {\bf c}'_1 = (0,0,0)$, ${\bf c}_2 = {\bf c}'_2 = (1,
\alpha,0)$, ${\bf c}_3 = (\alpha,0,1)$, and ${\bf c}'_3 =
(\alpha,\alpha+1,0)$, where $\alpha$ is a primitive element of the
field. It can be verified that $\dr({\bf c}_1,{\bf c}_2)=\dr({\bf
c}_2,{\bf c}_3)=\dr({\bf c}_3,{\bf c}_1)=2$ and $\dr({\bf c}'_1,{\bf
c}'_2)=\dr({\bf c}'_2,{\bf c}'_3)=\dr({\bf c}'_3,{\bf c}'_1)=2$.
However, $B_1({\bf c}_1) \cap B_1({\bf c}_2) \cap B_1({\bf c}_3) =
\{(\alpha+1,0,0)\}$, whereas $B_1({\bf c}'_1) \cap B_1({\bf c}'_2)
\cap B_1({\bf c}'_3) = \{(1,0,0), (0,\alpha+1,0),
(\alpha,\alpha,0)\}$. We remark that this is similar to the problem
of the intersection of three balls with Hamming radii discussed in
\cite[p. 58]{cohen_book97}, provided that the underlying field
$\mathrm{GF}(q^m)$ is not $\mathrm{GF}(2)$.

\section{Packing properties of rank metric
codes}\label{sec:packing}
Combining~(\ref{eq:singleton1})
and~(\ref{eq:singleton2}) and generalizing slightly to account for
nonlinear codes, we can show that the cardinality $K$ of a code $C$
over $\mathrm{GF}(q^m)$ with length $n$ and minimum rank distance
$\dr$ satisfies
\begin{equation}\label{eq:singleton3}
    K \leq \min \left\{q^{m(n-\dr+1)},q^{n(m-\dr+1)} \right\}.
\end{equation}
In this paper, we call the bound in (\ref{eq:singleton3}) the
Singleton bound\footnote{The Singleton bound in \cite{roth_it91} has
a different form since array codes are defined over base fields.}
for codes with the rank metric, and refer to codes that attain the
Singleton bound as maximum rank distance (MRD) codes.

For any given parameter set $n$, $m$, and $\dr$, explicit
construction for MRD codes exists. For $n\leq m$ and $\dr \leq n$,
generalized Gabidulin codes can be constructed. For $n
> m$ and $\dr \leq m$, an MRD code can be constructed by transposing
a generalized Gabidulin code of length $m$ and minimum rank distance
$\dr$ over GF$(q^n)$, and this code is not necessarily linear over
GF$(q^m)$. Although maximum distance separable (MDS) codes, which
attain the Singleton bound for the Hamming metric, exist only for
limited block length over any given field, MRD codes can be
constructed for any block length $n$ and minimum rank distance $\dr$
over arbitrary fields GF$(q^m)$. This has significant impact on the
packing properties of rank metric codes as explained below.

The sphere packing problem we consider is as follows: given a finite
field $\mathrm{GF}(q^m)$, length $n$, and radius $r$, what is the
maximum number of non-intersecting balls with radius $r$ that can be
packed into $\mathrm{GF}(q^m)^n$? The sphere packing problem is
equivalent to finding the maximum cardinality $A(q^m,n,d)$ of a code
over $\mathrm{GF}(q^m)$ with length $n$ and minimum distance $d \geq
2r+1$: the spheres of radius $r$ centered at the codewords of such a
code do not intersect one another. Furthermore, when these
non-intersecting spheres centered at all codewords cover the {\em
whole} space, the code is called a perfect code.

For the Hamming metric, although nontrivial perfect codes do exist,
the optimal solution to the sphere packing problem is not known for
all the parameter sets \cite{macwilliams_77}. In contrast, for rank
metric codes, although nontrivial perfect rank metric codes do not
exist \cite{babu_95, chen_mn96}, we show that MRD codes provide an
optimal solution to the sphere packing problem for any set of
parameters. For given $n$, $m$, and $r$, let us denote the maximum
cardinality among rank metric codes over $\mathrm{GF}(q^m)$ with
length $n$ and minimum distance $\dr = 2r+1$ as $A_{\mbox{\tiny
R}}(q^m,n,\dr)$. For $\dr > \min\{n,m\}$, $A_{\mbox{\tiny
R}}(q^m,n,\dr) = 1$. For $\dr \leq \min\{n,m\}$, $A_{\mbox{\tiny
R}}(q^m,n,\dr) = \min \left\{ q^{m(n-\dr+1)},
q^{n(m-\dr+1)}\right\}$. Note that the maximal cardinality is
achieved by MRD codes for all parameter sets. Hence, MRD codes admit
the optimal solutions to the sphere packing problem for rank metric
codes.

The performance of Hamming metric codes of large block length can be
studied in terms of asymptotic bounds on the relative minimum
distance in the limit of infinite block length. In this section, we
aim to derive the asymptotic form of $A_{\mbox{\tiny R}}(q^m,n,\dr)$
in the case where both block length and minimum rank distance go to
infinity. However, this cannot be achieved for finite $m$ since the
minimum rank distance is no greater than $m$. Thus, we consider the
case where $\lim_{n \rightarrow \infty}\frac{n}{m}=b$, where $b$ is
a constant.

Define $\delta \df \lim_{n\rightarrow\infty} \frac{d_{\mbox{\tiny
R}}}{n}$ and $a(\delta) \df \lim_{n\rightarrow
    \infty}\sup \left[ \frac{\log_{q^m}A_{\mbox{\tiny R}} (q^m,n,\lfloor\delta
    n\rfloor)}{n}\right]$, where $a(\delta)$ represents the maximum possible code rate of a
code which has relative minimum distance $\delta$ as its length goes
to infinity.
%
%
We can thus determine the maximum possible code rate $a(\delta)$ of
a code.

\begin{proposition}\label{prop:a(delta)}
For $0 \leq \delta \leq \min\{1,b^{-1}\}$, the existence of MRD
codes for all parameter sets implies that
\begin{equation}\label{eq:a(delta)}
    a(\delta) = \min \left\{ 1-\delta, 1-b\delta \right\}.
\end{equation}
\end{proposition}

\section{Covering properties of rank metric
codes}\label{sec:covering}

\subsection{The sphere covering problem}\label{sec:covering_intro}

In this section, we are interested in the sphere covering problem
for the rank metric. This problem can be stated as follows: given an
extension field $\mathrm{GF}(q^m)$, length $n$, and radius $\rho$,
we want to determine the minimum number of balls of rank radius
$\rho$ which cover $\mathrm{GF}(q^m)^n$ entirely. The sphere
covering problem is equivalent to finding the minimum cardinality
$K_{\mbox{\tiny R}}(q^m,n,\rho)$ of a code over $\mathrm{GF}(q^m)$
with length $n$ and rank covering radius $\rho$. We remark that if
$C$ is a code over $\mathrm{GF}(q^m)$ with length $n$ and covering
radius $\rho$, then its transpose code $C^T$ is a code over
$\mathrm{GF}(q^n)$ with length $m$ and the same covering radius.
Therefore, $K_{\mbox{\tiny R}}(q^m,n,\rho) = K_{\mbox{\tiny
R}}(q^n,m,\rho)$, and without loss of generality we shall assume $n
\leq m$ henceforth in this section.

We remark that $K_{\mbox{\tiny R}}(q^m,n,0) = q^{mn}$ and
$K_{\mbox{\tiny R}}(q^m,n,n) = 1$ for all $m$ and $n$. Hence we
assume $0< \rho <n$ throughout this section. Two bounds on
$K_{\mbox{\tiny R}}(q^m,n,\rho)$ can be easily derived.

\begin{proposition}\label{prop:obvious_bounds_K}
For a code over $\mathrm{GF}(q^m)$ with length $n$ and covering
radius $0 < \rho < n$, we have
\begin{equation}\label{eq:obvious_bounds_K}
    \left\lfloor \frac{q^{mn}}{V_{\rho}(q^m,n)} \right\rfloor + 1 \leq K_{\mbox{\tiny
    R}}(q^m,n,\rho) \leq q^{m(n-\rho)}.
\end{equation}
\end{proposition}

\begin{proof}
The lower bound is a straightforward generalization of the bound
given in \cite{vasantha_gs99}. Note that the only codes with
cardinality $\frac{q^{mn}}{V_{\rho}(q^m,n)}$ are perfect codes.
However, there are no nontrivial perfect codes for the rank metric
\cite{babu_95}. Therefore, $K_{\mbox{\tiny R}} (q^m,n,\rho) >
\frac{q^{mn}} {V_{\rho}(q^m,n)}$. The upper bound follows from $\rho
\leq n-k$ for any $(n,k)$ linear code \cite{cohen_book97}, and hence
any linear code with covering radius $\rho$ has cardinality $\leq
q^{m(n-\rho)}$.
\end{proof}

We refer to the lower bound in~(\ref{eq:obvious_bounds_K}) as the
sphere covering bound.

For a code over $\mathrm{GF}(q^m)$ with length $n$ and covering
radius $0 < \rho < n$, we have $K_{\mbox{\tiny R}}(q^m,n,\rho) \leq
K_{\mbox{{\tiny H}}}(q^m,n,\rho)$, where $K_{\mbox{{\tiny
H}}}(q^m,n,\rho)$ is the minimum cardinality of a (linear or
nonlinear) code over $\mathrm{GF}(q^m)$ with length $n$ and Hamming
covering radius $\rho$. This is because any code with Hamming
covering radius $\rho$ has rank covering radius $\leq \rho$. Since
$K_{\mbox{{\tiny H}}}(q^m,n,\rho) \leq q^{m(n-\rho)}$, this provides
a tighter bound than the one given in
Proposition~\ref{prop:obvious_bounds_K}.

\begin{lemma}\label{lemma:K_n_n'}
For all $m>0$ and nonnegative $n$, $n'$, $\rho$, and $\rho'$, we
have
\begin{equation}\label{eq:K_n_n'}
    \Kr(q^m,n+n',\rho+\rho') \leq \Kr(q^m,n,\rho)
    \Kr(q^m,n',\rho').
\end{equation}
In particular, we have
\begin{eqnarray}
    \label{eq:K_n+1_rho+1}
    \Kr(q^m,n+1,\rho+1) &\leq& \Kr(q^m,n,\rho),\\
    \label{eq:K_n+1_rho}
    \Kr(q^m,n+1,\rho) &\leq& q^m \Kr(q^m,n,\rho).
\end{eqnarray}
\end{lemma}

\begin{proof}
First,~(\ref{eq:K_n_n'}) follows directly from \cite[Proposition
4]{vasantha_gs99}. In particular, when $(n',\rho') = (1,1)$ and
$(n',\rho') = (1,0)$, we obtain~(\ref{eq:K_n+1_rho+1})
and~(\ref{eq:K_n+1_rho}) respectively.
\end{proof}

\subsection{Lower bounds for the sphere covering
problem}\label{sec:lower_bounds}

We will derive two nontrivial lower bounds on $K_{\mbox{\tiny
R}}(q^m,n,\rho)$. First, we adapt the bound given in \cite[Theorem
1]{cohen_it86}.

\begin{proposition}\label{prop:bound_K_cohen}
For all $q^m$, $n$, and $0 < \rho < n$, we have
\begin{equation}\label{eq:bound_K_cohen}
    K_{\mbox{\tiny R}}(q^m,n,\rho) \geq
    \left\lceil \frac{q^{mn}-A_{\mbox{\tiny R}}(q^m,n,2\rho+1)q^{\rho^2}{2\rho \brack
    \rho}} {V_{\rho}(q^m,n)-q^{\rho^2}{2\rho \brack \rho}} \right\rceil,
\end{equation}
provided that the denominator on the right hand side (RHS) is
positive.
\end{proposition}

\begin{proof}
Suppose $C$ is a code over $\mathrm{GF}(q^m)$ with length $n$ and
rank covering radius $\rho$, and let $C_0$ be a maximal subcode of
$C$ with minimum rank distance $d' \geq 2\rho+1$. If $d'>n$, we
choose $C_0$ to be a single codeword. $C_0$ thus covers
$|C_0|V_{\rho}(q^m,n)$ vectors. Define $C_1 = C \backslash C_0$,
($C_1$ is not empty, otherwise $C$ would be a nontrivial perfect
code) and for any ${\bf c}_1 \in C_1$, let $f({\bf c}_1)$ denote the
number of vectors covered by ${\bf c}_1$ which are not covered by
$C_0$. Since $C_0$ is maximal, there exists at least one codeword
${\bf c}_0 \in C_0$ such that $\dr({\bf c}_0,{\bf c}_1) \leq 2\rho$.
We have $f({\bf c}_1) \leq V_{\rho}(q^m,n) - q^{\rho^2}{2\rho \brack
\rho}$, where the equality corresponds to when there is only one
such ${\bf c}_0$ and $\dr({\bf c}_0,{\bf c}_1) = 2\rho$  by
Proposition~\ref{prop:inter_2_balls}. In that case,
Proposition~\ref{prop:v-a_problem} implies that there are
$q^{\rho^2}{2\rho \brack \rho}$ vectors covered by both ${\bf c}_0$
and ${\bf c}_1$. Thus, we have
\begin{eqnarray*}
    q^{mn}&\leq& |\mathcal{C}_0|V_{\rho}(q^m,n) + \sum_{{\bf c}_1 \in
    \mathcal{C}_1} f({\bf c}_1)\\
    &\leq& |\mathcal{C}_0|V_{\rho}(q^m,n) + (|\mathcal{C}|-|\mathcal{C}_0|) \left(
    V_{\rho}(q^m,n)
    - q^{\rho^2}{2\rho \brack \rho} \right)\\
    &=& |\mathcal{C}|\left( V_{\rho}(q^m,n) - q^{\rho^2}{2\rho \brack \rho}
    \right) + |\mathcal{C}_0|q^{\rho^2}{2\rho \brack \rho}.
\end{eqnarray*}
We have $|\mathcal{C}_0| \leq A_{\mbox{\tiny R}}(q^m,n,d') \leq
A_{\mbox{\tiny R}}(q^m,n,2\rho+1)$, and the result follows.
\end{proof}

We remark that $A_{\mbox{\tiny R}}(q^m,n,2\rho+1)$ is
$q^{m(n-2\rho)}$ if $2\rho+1 \leq n$, or $1$ otherwise. Next, we
obtain both sufficient and necessary conditions under which the
bound is nontrivial, i.e., when the denominator on the RHS of
(\ref{eq:bound_K_cohen}) is positive.

\begin{lemma}\label{lemma:K_cohen_sufficient}
The denominator on the RHS of~(\ref{eq:bound_K_cohen}) is positive
if $\rho(m+n-3\rho) \geq \sigma(q)$. Also, the denominator
in~(\ref{eq:bound_K_cohen}) is positive only if $m+n \geq 3\rho$.
\end{lemma}

\begin{proof}
We first prove the sufficient condition. We need to show
$V_{\rho}(q^m,n) > q^{\rho^2}{2\rho \brack \rho}$. By
Lemma~\ref{lemma:lower_bound_Vt}, $V_\rho(q^m,n) \geq
q^{\rho(m+n-\rho)}$. By \cite[Lemma 1]{gadouleau_it06}, we have
${2\rho \brack \rho} < q^{\rho^2+\sigma(q)}$. Therefore, the
denominator in~(\ref{eq:bound_K_cohen}) is positive if
$\rho(m+n-\rho) \geq 2\rho^2+\sigma(q)$.

We now prove the necessary condition. Note that $\alpha(n,\rho) \leq
q^{n\rho}$ and $\alpha(2\rho,\rho) \geq q^{2\rho^2-\tau(q)}$, where
$\tau(q) = \log_q \left( \frac{q^2}{q^2-1} \right)$ \cite[Lemma
2]{gadouleau_it06}. Now suppose $\rho(m+n-3\rho) < -\tau(q)$, then
$q^{\rho(m+n-3\rho) + \tau(q)} < 1$. This implies
$\frac{\alpha(n,\rho)} {\alpha(2\rho,\rho)} q^{\rho(m-\rho)} < 1$,
and hence ${n \brack \rho} q^{m\rho} < {2\rho \brack \rho}
q^{\rho^2}$. By \cite[Lemma 13]{gadouleau_it06}, we obtain
$V_\rho(q^m,n) \leq {n \brack \rho} q^{m\rho} < q^{\rho^2} {2\rho
\brack \rho}$. Therefore, $V_\rho(q^m,n) - q^{\rho^2} {2\rho \brack
\rho} > 0$ only if $\rho(m+n-3\rho) \geq -\tau(q)$. Finally, $0 <
\tau(q) < 1$ for $q \geq 2$ and hence $\rho(m+n-3\rho) \geq 0$.
\end{proof}

Before deriving the second nontrivial lower bound, we need the
following adaptation of \cite[Lemma 8]{vanwee_91}. Let $C$ be a code
with length $n$ and rank covering radius $\rho$ over
$\mathrm{GF}(q^m)$. We define $A \df \{{\bf x} \in
\mathrm{GF}(q^m)^n | \dr({\bf x},C) = \rho\}$.

\begin{lemma}\label{lemma:epsilon}
For ${\bf x} \in A \backslash Z$ and $0 < \rho < n$, we have
\begin{equation}\label{eq:E_C_epsilon}
    E_C(B_1({\bf x})) \geq \epsilon,
\end{equation}
where
\begin{equation}\label{eq:epsilon}
    \nonumber
    \epsilon \df \left\lceil
    \frac{(q^m-q^\rho)({n \brack 1} - {\rho \brack 1})}
    {q^{\rho}{\rho+1 \brack 1}} \right\rceil q^{\rho}{\rho+1 \brack 1}
    + (q^m-q^\rho) \left({\rho \brack 1} - {n \brack 1}\right).
\end{equation}
\end{lemma}

\begin{proof}
Since ${\bf x} \notin Z$, there is a unique ${\bf c}_0 \in C$ such
that $\dr({\bf x}, {\bf c_0}) = \rho$. By Proposition~\ref{prop:B_r}
we have $|B_{\rho}({\bf c_0}) \cap B_1({\bf x})| = 1+
(q^m-q^\rho){\rho \brack 1} + (q^\rho-1){n \brack 1}$. For any
codeword ${\bf c}_1 \in C$ satisfying $\dr({\bf x}, {\bf c}_1) =
\rho + 1$, by Proposition~\ref{prop:v-a_problem} we have
$|B_{\rho}({\bf c}_1) \cap B_1({\bf x})| = q^{\rho}{\rho+1 \brack
1}$. Finally, for all other codewords ${\bf c}_2 \in C$ at distance
$> \rho+1$ from ${\bf x}$, we have $|B_{\rho}({\bf c}_2) \cap
B_1({\bf x})| = 0$. Denoting $N \df |\{ {\bf c}_1 \in C | \dr({\bf
x},{\bf c}_1) = \rho+1 \}|$, we obtain
\begin{eqnarray*}
    E_C(B_1({\bf x})) &=& \sum_{{\bf c} \in C}|B_{\rho}({\bf c})
    \cap B_1({\bf x})| - |B_1({\bf x})|\\
    &=& (q^m-q^\rho){\rho \brack 1} + N q^{\rho}{\rho+1 \brack 1} -
    {n \brack 1}(q^m-q^\rho)\\
    &\equiv& (q^m-q^\rho)\left({\rho \brack 1} - {n \brack 1} \right)
    \mod \left( q^{\rho}{\rho+1 \brack 1} \right).
\end{eqnarray*}
The proof is completed by realizing that $(q^m-q^\rho)\left({\rho
\brack 1} - {n \brack 1} \right) < 0$, while $E_C(B_1({\bf x}))$ is
a non-negative integer.
\end{proof}

\begin{proposition}\label{prop:excess_bound}
If $\epsilon>0$, then
\begin{equation}\label{eq:excess_bound}
    K_{\mbox{\tiny R}}(q^m,n,\rho) \geq \left\lceil \frac{q^{mn}}
    {V_\rho(q^m,n) - \frac{\epsilon}{\delta} N_{\rho}(q^m,n)} \right\rceil,
\end{equation}
where $\delta \df V_1(q^m,n) - q^{\rho-1}{\rho \brack 1} -1 +
2\epsilon$.
\end{proposition}

The proof of Proposition~\ref{prop:excess_bound}, provided in
Appendix~\ref{app:prop:excess_bound}, uses the approach in the proof
of \cite[Theorem 6]{vanwee_91} and is based on the concept of excess
reviewed in Section~\ref{sec:covering_radius}. We remark that,
unlike the bound given in Proposition~\ref{prop:bound_K_cohen}, the
bound in Proposition~\ref{prop:excess_bound} is always applicable.
The lower bounds in~(\ref{eq:bound_K_cohen})
and~(\ref{eq:excess_bound}), when applicable, are at least as tight
as the sphere covering bound given in~(\ref{eq:obvious_bounds_K}).

\subsection{Upper bounds for the sphere covering problem}\label{sec:upper_bounds}
From the perspective of covering, the following lemma gives a
characterization of MRD codes in terms of ELS's.
\begin{lemma}\label{lemma:C+V}
Let $\mathcal{C}$ be an $(n,k)$ linear code over $\mathrm{GF}(q^m)$
($n \leq m$). $\mathcal{C}$ is an MRD code if and only if
$\mathcal{C} \oplus \mathcal{V} = \mathrm{GF}(q^m)^n$ for all
$\mathcal{V} \in E_{n-k}(q^m,n)$.
\end{lemma}
\begin{proof}
Suppose $\mathcal{C}$ is an $(n,k,n-k+1)$ MRD code. It is clear that
$\mathcal{C} \cap \mathcal{V} = \left\{{\mathbf 0}\right\}$ and
hence $\mathcal{C} \oplus \mathcal{V} = \mathrm{GF}(q^m)^n$ for all
$\mathcal{V} \in E_{n-k}(q^m,n)$.

Conversely, suppose $\mathcal{C} \oplus \mathcal{V} =
\mathrm{GF}(q^m)^n$ for all $\mathcal{V} \in E_{n-k}(q^m,n)$. Then
$\mathcal{C}$ does not contain any nonzero codeword of weight $\leq
n-k$, and hence its minimum distance is $n-k+1$.
\end{proof}

Let $\alpha_0, \alpha_1, \ldots,\alpha_{m+\rho-1} \in
\mathrm{GF}(q^{m+\rho})$ be a basis set of $\mathrm{GF}(q^{m+\rho})$
over $\mathrm{GF}(q)$, and let $\beta_0, \beta_1, \ldots,
\beta_{m-1}$ be a basis of $\mathrm{GF}(q^m)$ over $\mathrm{GF}(q)$.
We define the linear mapping $f$ between two vector spaces
$\mathrm{GF}(q^m)$ and $\mathfrak{S}_m \df \mathfrak{S}(\alpha_0,
\alpha_1, \ldots, \alpha_{m-1})$ by $f(\beta_i) = \alpha_i$ for $0
\leq i \leq m-1$. This can be generalized to $n$-dimensional
vectors, by applying $f$ componentwise. We thus define ${\bar f} :
\mathrm{GF}(q^m)^n \rightarrow \mathrm{GF}(q^{m+\rho})^n$ such that
for any ${\bf v} = (v_0,\ldots,v_{n-1})$, ${\bar f}({\bf v}) =
(f(v_0), \ldots, f(v_{m-1}))$. This function ${\bar f}$ is a linear
bijection from $\mathrm{GF}(q^m)^n$ to its image $\mathfrak{S}_m^n$.

\begin{lemma}\label{lemma:f(V)}
For all $r$ and any $\mathcal{V} \in E_r(q^m,n)$, ${\bar
f}(\mathcal{V}) \subset \mathcal{W}$, where $\mathcal{W} \in
E_r(q^{m+\rho},n)$.
\end{lemma}

\begin{proof}
We first show that ${\bar f}$ preserves the rank. Suppose ${\bf u}
\in \mathrm{GF}(q^m)^n$. Let us denote the matrix formed after
extending the coordinates of ${\bf u}$ with respect to the basis
$\{\beta_i\}$ as ${\bf U}$. The extension of ${\bar f}({\bf u})$
with respect to the basis $\{\alpha_i\}$ is given by $\bar{\bf U} =
\left(\begin{array}{c} {\bf U} \\ {\bf 0}
\end{array}\right)$. We thus have $\rk(\bar{\bf U}) = \rk({\bf U})$,
and $\rk({\bar f}({\bf u})) = \rk({\bf u})$.

Let $B = \{{\bf b}_i\}$ be a basis of $\mathcal{V} \in E_r(q^m,n)$
with vectors of rank one. Then for all $i$, ${\bar f}({\bf b}_i)$
has rank one and $\{{\bar f}({\bf b}_i)\}$ form a basis, and hence
${\bar f}(\mathcal{V}) \subset \mathcal{W}$, where $\mathcal{W} \in
E_r(q^{m+\rho},n)$ with $\{{\bar f}({\bf b}_i)\}$ as a basis.
\end{proof}

\begin{proposition}\label{prop:rho_MRD}
Let $\mathcal{C}$ be an $(n,n-\rho,\rho+1)$ MRD code with covering
radius $\rho$. Then the code ${\bar f}(\mathcal{C})$ is a code of
length $n$ over $\mathrm{GF}(q^{m+\rho})$ with cardinality
$q^{m(n-\rho)}$ and covering radius $\rho$ .
\end{proposition}

\begin{proof}
The other parameters for the code are obvious, and it suffices to
establish the covering radius. Let $\mathfrak{T}_\rho$ be a subspace
of $\mathrm{GF}(q^{m+\rho})$ with dimension $\rho$ such that
$\mathfrak{S}_m \oplus \mathfrak{T}_\rho = \mathrm{GF}(q^{m+\rho})$.
Any ${\bf u} \in \mathrm{GF}(q^{m+\rho})^n$ can be expressed as
${\bf u} = {\bf v} + {\bf w}$, where ${\bf v} \in \mathfrak{S}_m^n$
and ${\bf w} \in \mathfrak{T}_\rho^n$. Hence $\rk({\bf w}) \leq
\rho$, and ${\bf w} \in {\mathcal W}$ for some ${\mathcal W} \in
E_\rho(q^{m+\rho},n)$. By Lemma~\ref{lemma:C+V}, we can express
${\bf v}$ as ${\bf v} = {\bar f}({\bf c} + {\bf e}) = {\bar f}({\bf
c}) + {\bar f}({\bf e})$, where ${\bf c} \in \mathcal{C}$ and ${\bf
e} \in \mathcal{V}$, such that ${\bar f}(\mathcal{V}) \subset
{\mathcal W}$. Eventually, we have ${\bf u} = {\bar f}({\bf c}) +
{\bar f}({\bf e}) + {\bf w}$, where ${\bar f}({\bf e}) + {\bf w} \in
{\mathcal W}$, and thus $d({\bf u}, {\bar f}({\bf c})) \leq \rho$.
Thus $\bar{f}(\mathcal{C})$ has covering radius $\leq \rho$.
Finally, it is easy to verify that the covering radius of
$\bar{f}(\mathcal{C})$ is exactly $\rho$.
\end{proof}

\begin{corollary}\label{cor:bound_K}
We have
\begin{equation}\label{eq:bound_K_MRD1}
    \Kr(q^m,n,\rho) \leq q^{ \max\{m-\rho, n\} (n-\rho)}.
\end{equation}
\end{corollary}

\begin{proof}
We can construct an $(n,n-\rho)$  MRD code $\mathcal{C}$ over
$\mathrm{GF}(q^\mu)$ with covering radius $\rho$, where $\mu =
\max\{m-\rho,n\}$. By Proposition~\ref{prop:rho_MRD}, ${\bar
f}(\mathcal{C})$, where ${\bar f}$ maps $\mathrm{GF}(q^\mu)^n$ into
a subset of $\mathrm{GF}(q^m)^n$, has covering radius $\rho$. Note
that $|{\bar f}(\mathcal{C})| = |\mathcal{C}| = q^{\mu(n-\rho)}$.
\end{proof}

We can use the properties of $\Kr(q^m,n,\rho)$ in
Lemma~\ref{lemma:K_n_n'} in order to obtain two tighter bounds when
$\rho \geq m-n$.

\begin{proposition}\label{prop:bound_K_mixed}
Given fixed $m$, $n$, and $\rho$, for any $n\geq l>0$ and $(n_i,
\rho_i)$ for $0\leq i \leq l-1$ so that $0 < n_i \leq n$, $0 \leq
\rho_i \leq n_i$, and $n_i + \rho_i \leq m$ for all $i$, and
$\sum_{i=0}^{l-1} n_i = n$ and $\sum_{i=0}^{l-1} \rho_i = \rho$, we
have
\begin{equation}\label{eq:bound_K_mixed}
    \Kr(q^m,n,\rho) \leq \min_{\left\{(n_i, \rho_i): \,0\leq i \leq l-1\right\}} \left\{ q^{m(n-\rho) - \sum_i \rho_i(n_i-\rho_i)}
    \right\}.
\end{equation}
\end{proposition}

\begin{proof}
By Lemma~\ref{lemma:K_n_n'}, we have $\Kr(q^m,n,\rho) \leq \prod_i
\Kr(q^m,n_i,\rho_i)$ for all possible sequences $\{\rho_i\}$ and
$\{n_i\}$. For all $i$, we have $\Kr(q^m,n_i,\rho_i) \leq
q^{(m-\rho_i)(n-\rho_i)}$ by Corollary~\ref{cor:bound_K}, and hence
$\Kr(q^m,n,\rho) \leq q^{\sum_i(m-\rho_i)(n_i-\rho_i)} =
q^{m(n-\rho) - \sum_i \rho_i(n_i-\rho_i)}$.
\end{proof}

It is clear that the upper bound in~(\ref{eq:bound_K_mixed}) is
tighter than the upper bound in~(\ref{eq:obvious_bounds_K}). It can
also be shown that it is tighter than the bound
in~(\ref{eq:bound_K_MRD1}).

%
%

The following upper bound is an adaptation of \cite[Theorem
12.1.2]{cohen_book97}.

\begin{proposition}\label{prop:bound_K_12.1}
For any $m$, $n \leq m$, and $\rho < n$, there exists a code over
$\mathrm{GF}(q^m)$ of length $n$ and covering radius $\rho$ with
cardinality
\begin{equation}\label{eq:bound_K_12.1}
    \Kr(q^m,n,\rho) \leq \left\lfloor \frac{1} {1-
    \log_{q^{mn}}\left(q^{mn} - V_\rho(q^m,n) \right)}
    \right\rfloor + 1.
\end{equation}
\end{proposition}

Our proof, given in Appendix~\ref{app:prop:bound_K_12.1}, adopts the
approach used to prove \cite[Theorem 12.1.2]{cohen_book97}.

%
%

\begin{proposition}\label{prop:JSL}
For all $m$, $n \leq m$, $\rho < n$, we have
\begin{equation}\label{eq:bound_JSL}
    \Kr(q^m,n,\rho) \leq \frac{q^{mn}}{V_\rho(q^m,n)}\left[1 +
    \ln(V_\rho(q^m,n))\right].
\end{equation}
\end{proposition}

\begin{proof}
Consider the square $0$-$1$ matrix ${\bf A}$ of order $q^{mn}$,
where each row and each column represents a vector in
$\mathrm{GF}(q^m)^n$. Set $a_{i,j} = 1$ if and only if the sphere
with rank radius $\rho$ centered at vector $i$ covers the vector
$j$. There are thus exactly $V_\rho(q^m,n)$ ones in each row and
each column of ${\bf A}$. Note that any $q^{mn} \times K$ submatrix
${\bf C}$ of ${\bf A}$ with no all-zeros rows represents a code with
cardinality $K$ and covering radius $\rho$. Applying the
Johnston-Stein-Lov\'asz theorem \cite[Theorem 12.2.1]{cohen_book97}
to ${\bf A}$, we can find such a submatrix with $K \leq
\frac{1}{V_\rho(q^m,n)}\left[q^{mn} + q^{mn}
    \ln(V_\rho(q^m,n))\right].$
\end{proof}

The tightest bounds on $\Kr(q^m,n,\rho)$ known so far are given in
Table~\ref{table:bounds} for $q=2$, $2 \leq m \leq 7$, $2 \leq n
\leq m$, and $1 \leq \rho \leq 6$.

%
%
%
%
%
%

\subsection{Covering properties of linear rank metric codes}
\label{sec:covering_linear}

For a linear code with given covering radius, the sphere covering
bound also implies a lower bound on its dimension.

\begin{proposition}\label{prop:bound_k_rho}
An $(n,k)$ linear code over $\mathrm{GF}(q^m)$ with rank covering
radius $\rho$ satisfies
\begin{equation}\label{eq:bound_k_rho}
  \left\lfloor n- \rho - \frac{\rho(n-\rho)+\sigma(q)}{m} \right\rfloor
  + 1 \leq k \leq n- \rho.
\end{equation}
\end{proposition}

\begin{proof}
The upper bound directly follows the upper bound
in~(\ref{eq:obvious_bounds_K}). We now prove the lower bound. By the
sphere covering bound, we have $q^{mk}
> \frac{q^{mn}}{V_\rho(q^m,n)}$. However, by
Lemma~\ref{lemma:lower_bound_Vt} we have $V_\rho(q^m,n) <
q^{\rho(m+n-\rho) + \sigma(q)}$ and hence $q^{mk} > q^{mn -
\rho(m+n-\rho) - \sigma(q)}$.
\end{proof}

We do not adapt the bounds in~(\ref{eq:bound_K_cohen})
and~(\ref{eq:excess_bound}) as their advantage over the lower bound
in~(\ref{eq:bound_k_rho}) is not significant.
Table~\ref{table:linear_bounds} lists the values of the bounds in
Proposition~\ref{prop:bound_k_rho} on the dimension of a linear code
with given covering radius for $4 \leq m \leq 8$ and $4 \leq n \leq
m$. Note that only $1 < \rho < n-1$ are considered in
Table~\ref{table:linear_bounds} since, as shown below, the dimension
of a linear code with given covering radius can be completely
determined when $\rho \in \{0,1,n-1,n\}$.

\begin{proposition}\label{prop:maximal_k}
Let $\mathcal{C}$ be an $(n,k)$ linear code over $\mathrm{GF}(q^m)$
($n \leq m$) with rank covering radius $\rho$. Then $k=n-\rho$ if
$\rho \in \{0,1,n-1,n\}$ or $\rho(n-\rho) \leq m - \sigma(q)$, or if
$\mathcal{C}$ is a generalized Gabidulin code or an \ELS{}.
\end{proposition}

\begin{proof}
The cases $\rho \in \{0,n-1,n\}$ are straightforward. In all other
cases, since $k \leq n-\rho$ by Proposition~\ref{prop:bound_k_rho},
it suffices to prove that $k \geq n-\rho$. First, suppose $\rho =
1$, then $k$ satisfies $q^{mk} > \frac{q^{mn}}{V_1(q^m,n)}$ by the
sphere covering bound. However, $V_1(q^m,n) < q^{m+n} \leq q^{2m}$,
and hence $k > n-2$. Second, if $\rho(n-\rho) \leq m - \sigma(q)$,
then $0< \frac{1}{m}\left( \rho(n-\rho) + \sigma(q) \right) \leq 1$
and $k \geq n-\rho $ by Proposition~\ref{prop:bound_k_rho}. Third,
if $\mathcal{C}$ is an $(n,k,n-k+1)$ generalized Gabidulin code with
$k<n$, then there exists an $(n,k+1,n-k)$ generalized Gabidulin code
$\mathcal{C}'$ such that $\mathcal{C} \subset \mathcal{C}'$. We have
$\rho \geq \dr(\mathcal{C}') = n-k$, as noted in
Section~\ref{sec:covering_radius}, and hence $k \geq n - \rho$. The
case $k=n$ is straightforward. Finally, if $\mathcal{C}$ is an
\ELS{} of dimension $k$, then for all ${\bf x}$ with rank $n$ and
for any ${\bf c} \in \mathcal{C}$, $\dr({\bf x}, {\bf c}) \geq
\rk({\bf x}) - \rk({\bf c}) \geq n-k$.
\end{proof}

A similar argument can be used to bound the covering radius of the
cartesian products of generalized Gabidulin codes.

\begin{corollary}\label{cor:radius_cartesian}
Let $\mathcal{G}$ be an $(n,k,d_{\mbox{\tiny R}})$ generalized
Gabidulin code $(n \leq m)$, and let $\mathcal{G}^l$ be the code
obtained by $l$ cartesian products of $\mathcal{G}$ for $l \geq 1$.
Then the rank covering radius of $\mathcal{G}^l$ satisfies
$\rho(\mathcal{G}^l) \geq \dr - 1$.
\end{corollary}

Note that when $n=m$, $\mathcal{G}^l$ is a maximal code, and hence
Corollary~\ref{cor:radius_cartesian} can be further strengthened.

\begin{corollary}\label{cor:radius_cartesian_m}
Let $\mathcal{G}$ be an $(m,k,\dr)$ generalized Gabidulin code over
$\mathrm{GF}(q^m)$, and let $\mathcal{G}^l$ be the code obtained by
$l$ cartesian products of $\mathcal{G}$. Then $\rho(\mathcal{G}^l) =
\dr-1$.
\end{corollary}

The tightest bounds for the dimension of linear codes with given
covering radius known so far are given in
Table~\ref{table:linear_bounds} for $q=2$, $4 \leq m \leq 8$, $4
\leq n \leq m$, and $2 \leq \rho \leq 6$.

\subsection{Asymptotic covering properties}\label{sec:asym_covering}
Table~\ref{table:bounds} provides solutions to the sphere covering
problem for only small values of $m$, $n$, and $\rho$.
Next, we study the asymptotic covering properties when both block
length and minimum rank distance go to infinity. As in
Section~\ref{sec:packing}, we consider the case where $\lim_{n
\rightarrow \infty}\frac{n}{m}=b$, where $b$ is a constant. In other
words, these asymptotic covering properties provide insights on the
covering properties of long rank metric codes over large fields.

The asymptotic form of the bounds in~(\ref{eq:lower_bound_Vt}) are
given in the lemma below.
\begin{lemma}\label{lemma:v(delta)}
For $0 \leq \delta \leq \min\{1,b^{-1}\}$, $v(\delta) \df \lim_{n
\rightarrow \infty} \left[ \frac{\log_{q^m} V_{\lfloor\delta
n\rfloor}(q^m,n)}{n}\right] = \delta(1+b-b\delta)$.
\end{lemma}

\begin{proof}
By Lemma~\ref{lemma:lower_bound_Vt}, we have $q^{\dr(m+n-\dr)} \leq
V_{\dr}(q^m,n) < q^{\dr(m+n-\dr) + \sigma(q)}$. Taking the logarithm
and dividing by $n$, this becomes $\delta(1+b-b\delta) \leq
\log_{q^m}( V_{\lfloor\delta n\rfloor} (q^m,n) )/n < \delta
(1+b-b\delta) + \frac{\sigma(q)}{mn}$. The proof is concluded by
taking the limit when $n$ tends to infinity.
\end{proof}

Define $r \df \frac{\rho}{n}$ and $k(r) = \lim_{n \rightarrow
\infty} \inf \left[ \frac{\log_{q^m}K_{\mbox{\tiny
R}}(q^m,n,\rho)}{n} \right]$. The bounds
in~(\ref{eq:obvious_bounds_K}) and~(\ref{eq:bound_JSL}) together
solve the asymptotic sphere covering problem.

\begin{theorem}\label{th:asymptotic_covering}
For all $b$ and $r$, we have
\begin{equation}\label{eq:asymptotic_covering}
    k(r) = (1-r)(1-br).
\end{equation}
\end{theorem}

\begin{proof}
By Lemma~\ref{lemma:v(delta)} the sphere covering bound
in~(\ref{eq:obvious_bounds_K}) asymptotically becomes $k(r) \geq
(1-r) (1-br)$. Also, from the bound in~(\ref{eq:bound_JSL}), we have
\begin{eqnarray*}
    \Kr(q^m,n,\rho) &\leq& \frac{q^{mn}}{V_\rho(q^m,n)}
    \left[ 1 + \ln(V_n(q^m,n)) \right]\\
    &\leq& \frac{q^{mn}}{V_\rho(q^m,n)}
    \left[ 1 + mn\ln(q) \right]\\
    \log_{q^{mn}} \Kr(q^m,n,\rho) &\leq& \log_{q^{mn}}
    \frac{q^{mn}}{V_\rho(q^m,n)} + O((mn)^{-1} \ln(mn)).
\end{eqnarray*}
By Lemma~\ref{lemma:v(delta)}, this asymptotically becomes $k(r)
\leq (1-r)(1-br)$. Note that although we assume $n \leq m$ above for
convenience, both bounds in~(\ref{eq:obvious_bounds_K})
and~(\ref{eq:bound_JSL}) hold for any values of $m$ and $n$.
\end{proof}



\section{MacWilliams identity}\label{sec:macwilliams}

For all ${\bf v} \in \mathrm{GF}(q^m)^n$ with rank weight $r$, the
rank weight function of ${\bf v}$ is defined as $f_{\mbox{\tiny
R}}({\bf v}) = y^{r}x^{n-r}$. Let ${\mathcal{C}}$ be a code of
length $n$ over $\mathrm{GF}(q^m)$. Suppose there are $A_i$
codewords in $\mathcal{C}$ with rank weight $i$ ($0 \leq i \leq n$),
then the rank weight enumerator of ${\mathcal{C}}$, denoted as
$W^{\mbox{\tiny R}}_{\mathcal{C}}(x,y)$, is defined to be
$$
    W^{\mbox{\tiny R}}_{\mathcal{C}}(x,y) \df \sum_{{\bf v} \in {\mathcal{C}}}f_{\mbox{\tiny R}}({\bf v}) = \sum_{i=0}^n A_i y^{i}x^{n-i}.
$$

\subsection{$q$-product of homogeneous polynomials}
\label{sec:star_prod_q-derivative}

Let $a(x,y;m) = \sum_{i=0}^r a_{i}(m) y^i x^{r-i}$ and $b(x,y;m) =
\sum_{j=0}^s b_{j}(m) y^j x^{s-j}$ be two homogeneous polynomials in
$x$ and $y$ of degrees $r$ and $s$ respectively with coefficients
$a_i(m)$ and $b_j(m)$ respectively. $a_i(m)$ and $b_j(m)$ for $i,j
\geq 0$ in turn are real functions of $m$, and are assumed to be
zero unless otherwise specified.

\begin{definition}[$q$-product]\label{def:star_prod}
The {\em $q$-product} of $a(x,y;m)$ and $b(x,y;m)$ is defined to be
the homogeneous polynomial of degree $(r+s)$ $c(x,y;m) \df a(x,y;m)
* b(x,y;m) = \sum_{u=0}^{r+s} c_{u}(m) y^u x^{r+s-u}$, with
\begin{equation}
    c_{u}(m) = \sum_{i=0}^u q^{is} a_{i}(m) b_{u-i}(m-i).
\label{eq:q-product}\end{equation}

We shall denote the $q$-product by $*$ henceforth. For $n \geq 0$
the $n$-th $q$-power of $a(x,y;m)$ is defined recursively:
$a(x,y;m)^{[0]} = 1$ and $a(x,y;m)^{[n]} =
a(x,y;m)^{[n-1]}*a(x,y;m)$ for $n \geq 1$.
\end{definition}

We provide some examples to illustrate the concept. It is easy to
verify that $x * y = yx$, $y * x = qyx$, $yx * x = q yx^2$, and $yx
* (q^m-1)y = (q^m-q)y^2x$. Note that $x * y \ne y * x$. It is easy
to verify that the $q$-product is neither commutative nor
distributive in general. However, it is commutative and distributive
in some special cases as described below.

\begin{lemma}\label{lemma:properties_star_prod}
Suppose $a(x,y;m) = a$ is a constant independent from $m$, then
$a(x,y;m)*b(x,y;m) = b(x,y;m)*a(x,y;m) = ab(x,y;m)$. Also, if
$\deg[c(x,y;m)] = \deg[a(x,y;m)]$, then
$[a(x,y;m)+c(x,y;m)]*b(x,y;m) = a(x,y;m)*b(x,y;m) +
c(x,y;m)*b(x,y;m)$, and $b(x,y;m) * [a(x,y;m)+c(x,y;m)] =
b(x,y;m)*a(x,y;m) + b(x,y;m)*c(x,y;m)$.
\end{lemma}

The homogeneous polynomials $a_l(x,y;m) \df [x+(q^m-1)y]^{[l]}$ and
$b_l(x,y;m) \df (x-y)^{[l]}$ are very important to our derivations
below. The following lemma provides the analytical expressions of
$a_l(x,y;m)$ and $b_l(x,y;m)$.

\begin{lemma}\label{lemma:special_prod}
For $i \geq 0$, $\sigma_i \df \frac{i(i-1)}{2}$. For $l\geq 0$, we
have $y^{[l]} = q^{\sigma_l}y^l$ and $x^{[l]} = x^l$. Furthermore,
\begin{eqnarray}
    \label{eq:x+y^s}
    a_l(x,y;m) = \sum_{u=0}^l {l \brack u} \alpha(m,u) y^u
    x^{l-u},\\
    \label{eq:x-y^r}
    b_l(x,y;m) = \sum_{u=0}^l {l \brack u} (-1)^u q^{\sigma_u} y^u x^{l-u}.
\end{eqnarray}
\end{lemma}

Note that $a_l(x,y;m)$ is the rank weight enumerator of
$\mathrm{GF}(q^m)^l$. The proof of Lemma~\ref{lemma:special_prod} is
given in Appendix~\ref{app:lemma:special_prod}.

\begin{definition}[$q$-transform]\label{def:star_transform}
We define the {\em $q$-transform} of $a(x,y;m)= \sum_{i=0}^r
a_{i}(m) y^i x^{r-i}$ as the homogeneous polynomial $\bar{a}(x,y;m)=
\sum_{i=0}^r a_{i}(m) y^{[i]}* x^{[r-i]}$.
\end{definition}

\begin{definition}[$q$-derivative \cite{gasper_book04}]
\label{def:q-derivative} For $q \geq 2$, the $q$-derivative at $x
\neq 0$ of a real-valued function $f(x)$ is defined as
$$
    f^{(1)}(x) \df \frac{f(qx)-f(x)}{(q-1)x}.
$$
\end{definition}

For any real number $a$, $[f(x)+ag(x)]^{(1)} = f^{(1)}(x) +
ag^{(1)}(x)$ for $x \neq 0$. For $\nu \geq 0$, we shall denote the
$\nu$-th $q$-derivative (with respect to $x$) of $f(x,y)$ as
$f^{(\nu)}(x,y)$. The $0$-th $q$-derivative of $f(x,y)$ is defined
to be $f(x,y)$ itself.

\begin{lemma}\label{lemma:special_q-d}
For $0 \leq \nu \leq l$, $(x^l)^{(\nu)} = \beta(l,\nu)x^{l-\nu}$.
The $\nu$-th $q$-derivative of $f(x,y)= \sum_{i=0}^r f_i y^i
x^{r-i}$ is given by $f^{(\nu)}(x,y) = \sum_{i=0}^{r-\nu} f_i
\beta(i,\nu) y^i x^{r-i-\nu}$. Also,
\begin{eqnarray}
    \label{eq:al_nu}
    a_l^{(\nu)}(x,y;m) &=& \beta(l,\nu) a_{l-\nu}(x,y;m)\\
    \label{eq:bl_nu}
    b_l^{(\nu)}(x,y;m) &=& \beta(l,\nu) b_{l-\nu}(x,y;m).
\end{eqnarray}
\end{lemma}

The proof of Lemma~\ref{lemma:special_q-d} is given in
Appendix~\ref{app:lemma:special_q-d}.

\begin{lemma}[Leibniz rule for the $q$-derivative]\label{lemma:Leibniz_x}
For two homogeneous polynomials $f(x,y) = \sum_{i=0}^r f_i y^i
x^{r-i}$ and $g(x,y) = \sum_{j=0}^s g_j y^j x^{s-j}$ with degrees
$r$ and $s$ respectively, the $\nu$-th ($\nu \geq 0$) $q$-derivative
of their $q$-product is given by
\begin{equation}
    \label{eq:leibniz_nu}
    \left[f(x,y)*g(x,y)\right]^{(\nu)} = \sum_{l=0}^{\nu} {\nu \brack l}
    q^{(\nu-l)(r-l)} f^{(l)}(x,y)*g^{(\nu-l)}(x,y).
\end{equation}
\end{lemma}

The proof of Lemma~\ref{lemma:Leibniz_x} is given in
Appendix~\ref{app:lemma:leibniz_x}. Lemma~\ref{lemma:Leibniz_x}
gives the $\nu$-th $q$-derivative of $q$-products of homogeneous
polynomials.

The $q^{-1}$-derivative is similar to the $q$-derivative.
\begin{definition}[$q^{-1}$-derivative]\label{def:q-1_derivative}
For $q \geq 2$, the $q$-derivative at $y \neq 0$ of a real-valued
function $g(y)$ is defined as
$$
    g^{\{1\}}(y) \df \frac{g(q^{-1}y) - g(y)}{(q^{-1} - 1)y}.
$$
\end{definition}

For $\nu \geq 0$, we shall denote the $\nu$-th $q^{-1}$-derivative
(with respect to $y$) of $g(x,y)$ as $g^{\{\nu\}}(x,y)$. The $0$-th
$q^{-1}$-derivative of $g(x,y)$ is defined to be $g(x,y)$ itself.

\begin{lemma}\label{lemma:special_q-1_d}
For $0 \leq \nu \leq l$, the $\nu$-th $q^{-1}$-derivative of $y^l$
is $(y^l)^{\{ \nu \}} = q^{\nu(1-n) + \sigma_\nu} \beta(l,\nu)
y^{l-\nu}$. Also,
\begin{eqnarray}
    \label{eq:al_nu_q-1}
    a_l^{\{ \nu \}}(x,y;m) &=& \beta(l,\nu) q^{-\sigma_\nu} \alpha(m,\nu)
    a_{l-\nu}(x,y;m-\nu)\\
    \label{eq:bl_nu_q-1}
    b_l^{\{ \nu \}}(x,y;m) &=& (-1)^{\nu}\beta(l,\nu)b_{l-\nu}(x,y;m).
\end{eqnarray}
\end{lemma}

The proof of Lemma~\ref{lemma:special_q-1_d} is given in
Appendix~\ref{app:lemma:special_q-1_d}.

\begin{lemma}[Leibniz rule for the $q^{-1}$-derivative]\label{lemma:Leibniz_y}
For two homogeneous polynomials $f(x,y;m) = \sum_{i=0}^r f_i y^i
x^{r-i}$ and $g(x,y;m) = \sum_{j=0}^s g_j y^j x^{s-j}$ with degrees
$r$ and $s$ respectively, the $\nu$-th ($\nu \geq 0$)
$q^{-1}$-derivative of their $q$-product is given by
\begin{equation}\label{eq:leibniz_y_nu}
    [f(x,y;m)*g(x,y;m)]^{\{ \nu \}} = \sum_{l=0}^{\nu} {\nu \brack l}
    q^{l(s-\nu+l)} f^{\{ l \}}(x,y;m)*g^{\{ \nu-l \}}(x,y;m-l).
\end{equation}
\end{lemma}

The proof of Lemma~\ref{lemma:Leibniz_y} can be found in
Appendix~\ref{app:lemma:leibniz_y}.

\subsection{The dual of a vector}\label{sec:dual_v_rank}

For all ${\bf u}, {\bf v} \in \mathrm{GF}(q^m)^n$, let ${\bf u}
\cdot {\bf v}$ denote the standard inner product of ${\bf u}$ and
${\bf v}$. For any linear subspace $\mathcal{L} \subseteq
\mathrm{GF}(q^m)^n$, $\mathcal{L}^\perp \df \{ {\bf u} \in
\mathrm{GF}(q^m)^n | {\bf u} \cdot {\bf v} = 0  \,  \forall {\bf v}
\in \mathcal{L}\}$ is referred to as the dual of $\mathcal{L}$.

As an important step toward our main result, we derive the rank
weight enumerator of $\vspan{{\bf v}}^\perp$, where ${\bf v} \in
\mathrm{GF}(q^m)^{n}$ is an arbitrary vector and $\vspan{{\bf v}}\df
\left\{a{\bf v}: a \in \mathrm{GF}(q^m)\right\}$. Note that
$\vspan{{\bf v}}$ can be viewed as an $(n,1)$ linear code over
$\mathrm{GF}(q^m)$ with a generator matrix ${\bf v}$. It is
remarkable that the rank weight enumerator of $\vspan{{\bf
v}}^\perp$ depends on only the rank of ${\bf v}$.

Berger \cite{berger_it03} has determined that the linear isometries
for the rank distance are given by the scalar multiplication by a
non-zero element of $\mathrm{GF}(q^m)$, and multiplication on the
right by an nonsingular matrix ${\bf B} \in \mathrm{GF}(q)^{n \times
n}$. We say that two codes $C$ and $C'$ are rank-equivalent if there
exists a linear isometry $f$ for the rank distance such that $f(C) =
C'$.

\begin{lemma}\label{lemma:rk_dual_vector}
Suppose ${\bf v}$ has rank $r \geq 1$, Then $\mathcal{L} =
\vspan{{\bf v}}^\perp$ is rank-equivalent to $\mathcal{C} \times
\mathrm{GF}(q^m)^{n-r}$, where $\mathcal{C}$ is an $(r,r-1,2)$ MRD
code and $\times$ denotes cartesian product.
\end{lemma}

\begin{proof}
We can express ${\bf v}$ as ${\bf v} = \bar{\bf v} {\bf B}$, where
$\bar{\bf v} = (v_0,\ldots,v_{r-1},0\ldots,0)$ has rank $r$, and
${\bf B} \in \mathrm{GF}(q)^{n \times n}$ has full rank. Remark that
$\bar{\bf v}$ is the parity-check of the code $\mathcal{C} \times
\mathrm{GF}(q^m)^{n-r}$, where $\mathcal{C} =
\vspan{(v_0,\ldots,v_{r-1})}^\perp$ is an $(r,r-1,2)$ MRD code. It
can be easily checked that ${\bf u} \in \mathcal{L}$ if and only if
$\bar{\bf u} \df {\bf u}{\bf B}^T \in \vspan{\bar{\bf v}}^\perp$.
Therefore, $\vspan{\bar{\bf v}}^\perp = \mathcal{L} {\bf B}^T$, and
hence $\mathcal{L}$ is rank-equivalent to $\vspan{\bar{\bf v}}^\perp
= \mathcal{C} \times \mathrm{GF}(q^m)^{n-r}$.
\end{proof}

We hence derive the rank weight enumerator of an $(r,r-1,2)$ MRD
code. Note that the rank weight distribution of linear Class-I MRD
codes has been derived in \cite{delsarte_78,gabidulin_pit0185}.
However, we will use our results to give an alternative derivation
of the rank weight distribution of linear Class-I MRD codes later,
and thus we shall not use the result in
\cite{delsarte_78,gabidulin_pit0185} here.

\begin{proposition}\label{prop:trivial_MRD}
Suppose ${\bf v}_r \in \mathrm{GF}(q^m)^r$ has rank $r$ ($0\leq r
\leq m$). The rank weight enumerator of $\mathcal{L}_r = \vspan{{\bf
v}}^{\perp}$ depends on only $r$ and is given by
\begin{equation}
    W_{\mathcal{L}_r}^{\mbox{\tiny R}}(x,y) =
     q^{-m}\left\{ \left[x+(q^m-1)y \right]^{[r]} + (q^m-1)(x-y)^{[r]}\right\}.
\label{eq:trivial_MRD}\end{equation}
\end{proposition}

\begin{proof}
We first prove that the number of vectors with rank $r$ in
$\mathcal{L}_r$, denoted as $A_{r,r}$, depends only on $r$ and is
given by
\begin{equation}\label{eq:App}
    A_{r,r} = q^{-m}[\alpha(m,r) + (q^m-1)(-1)^r q^{\sigma_r}]
\end{equation}
by induction on $r$ ($r \geq 1$). Eq.~(\ref{eq:App}) clearly holds
for $r=1$. Suppose Eq.~(\ref{eq:App}) holds for $r = {\bar r}-1$.

We consider all the vectors ${\bf u} = (u_0,\ldots,u_{{\bar r}-1})
\in \mathcal{L}_{\bar r}$ such that the first ${\bar r}-1$
coordinates of ${\bf u}$ are linearly independent. Remark that
$u_{{\bar r}-1} = -v_{{\bar r}-1}^{-1} \sum_{i=0}^{{\bar r}-2}
u_iv_i$ is completely determined by $u_0,\ldots,u_{{\bar r}-2}$.
Thus there are $N_{{\bar r}-1}(q^m,{\bar r}-1) = \alpha(m,{\bar r}
-1)$ such vectors ${\bf u}$. Among these vectors, we will enumerate
the vectors ${\bf t}$ whose last coordinate is a linear combination
of the first ${\bar r}-1$ coordinates, i.e., ${\bf t} =
(t_0,\ldots,t_{{\bar r}-2},\sum_{i=0}^{{\bar r}-2} a_it_i) \in
\mathcal{L}_{\bar r}$ where $a_i \in \mathrm{GF}(q)$ for $0 \leq i
\leq {\bar r}-2$.

Remark that ${\bf t} \in \mathcal{L}_{\bar r}$ if and only if
$(t_0,\ldots,t_{{\bar r}-2}) \cdot (v_0 + a_0 v_{{\bar r}-1},
\ldots, v_{{\bar r}-2} + a_{{\bar r}-2} v_{{\bar r}-1}) = 0$. It is
easy to check that ${\bf v}({\bf a}) = (v_0 + a_0 v_{{\bar r}-1},
\ldots, v_{{\bar r}-2} + a_{{\bar r}-2} v_{{\bar r}-1})$ has rank
${\bar r}-1$. Therefore, if $a_0, \ldots, a_{{\bar r}-2}$ are fixed,
then there are $A_{{\bar r}-1,{\bar r}-1}$ such vectors ${\bf t}$.
Also, suppose $\sum_{i=0}^{{\bar r}-2} t_i v_i + v_{{\bar r}
-1}\sum_{i=0}^{{\bar r}-2} b_it_i = 0$. Hence $\sum_{i=0}^{{\bar r}
-2} (a_i -b_i)t_i=0$, which implies ${\bf a} = {\bf b}$ since
$t_i$'s are linearly independent. That is, $\vspan{{\bf v}({\bf
a})}^\perp \cap \vspan{{\bf v}({\bf b})}^\perp = \left\{{\bf
0}\right\}$ if ${\bf a} \neq {\bf b}$. We conclude that there are
$q^{{\bar r}-1} A_{{\bar r}-1,{\bar r}-1}$ vectors ${\bf t}$.
Therefore, $A_{{\bar r}, {\bar r}} = \alpha(m,{\bar r}-1) - q^{{\bar
r}-1} A_{{\bar r}-1,{\bar r}-1} = q^{-m} [\alpha(m,{\bar r}) +
(q^m-1)(-1)^{\bar r} q^{\sigma_{\bar r}}]$.

Denote the number of vectors with rank $p$ in $\mathcal{L}_r$ as
$A_{r,p}$. We have $A_{r,p} = {r \brack p}A_{p,p}$
\cite{gabidulin_pit0185}, and hence $A_{r,p} = {r \brack p} q^{-m}
[\alpha(m,p) + (q^m-1)(-1)^p q^{\sigma_p}]$. Thus, $W^{\mbox{\tiny
R}}_{\mathcal{L}_r}(x,y) = \sum_{p=0}^r A_{r,p} x^{r-p}y^p = q^{-m}
\Big\{ \left[x+(q^m-1)y \right]^{[r]} + (q^m-1)(x-y)^{[r]} \Big\}$.
\end{proof}

We comment that Proposition~\ref{prop:trivial_MRD} in fact provides
the rank weight distribution of any $(r,r-1,2)$ MRD code.
%

\begin{lemma}\label{lemma:rk_Asu1}
Let $\mathcal{C}_0 \subseteq \mathrm{GF}(q^m)^r$ be a linear code
with rank weight enumerator $W_{\mathcal{C}_0}^{\mbox{\tiny
R}}(x,y)$, and for $s \geq 0$, let $W_{\mathcal{C}_s}^{\mbox{\tiny
R}}(x,y)$ be the rank weight enumerator of ${\mathcal{C}_s} \df
\mathcal{C}_0 \times \mathrm{GF}(q^m)^s$. Then
$W_{\mathcal{C}_s}^{\mbox{\tiny R}}(x,y)$ only depends on $s$ and is
given by
\begin{equation}\label{eq:rk_Asu}
    W_{\mathcal{C}_s}^{\mbox{\tiny R}}(x,y) = W_{\mathcal{C}_0}^{\mbox{\tiny R}}(x,y) *
    \left[x+(q^m-1)y\right]^{[s]}.
\end{equation}
\end{lemma}

\begin{proof}
For $s \geq 0$, denote $W_{\mathcal{C}_s}^{\mbox{\tiny R}}(x,y) =
\sum_{u=0}^{r+s} B_{s,u} y^u x^{r+s-u}$. We will prove that
\begin{equation}\label{eq:Bsu}
    B_{s,u} = \sum_{i=0}^u q^{is} B_{0,i} {s \brack u-i}
    \alpha(m-i,u-i)
\end{equation}
by induction on $s$. Eq.~(\ref{eq:Bsu}) clearly holds for $s=0$. Now
assume (\ref{eq:Bsu}) holds for $s = {\bar s}-1$. For any ${\bf
x}_{\bar s} = (x_0,\ldots,x_{r+{\bar s}-1}) \in \mathcal{C}_{\bar
s}$, we define ${\bf x}_{{\bar s}-1} = (x_0,\ldots,x_{r+{\bar s}-2})
\in \mathcal{C}_{{\bar s}-1}$. Then $\rk({\bf x}_{\bar s}) = u$ if
and only if either $\rk({\bf x}_{{\bar s}-1}) = u$ and $x_{r+{\bar
s}-1} \in \mathfrak{S}({\bf x}_{{\bar s}-1})$ or $\rk({\bf x}_{{\bar
s}-1}) = u-1$ and $x_{r+{\bar s}-1} \notin \mathfrak{S} ({\bf
x}_{{\bar s}-1})$. This implies $B_{{\bar s},u} = q^u B_{{\bar
s}-1,u} + (q^m-q^{u-1})B_{{\bar s}-1,u-1}= \sum_{i=0}^u q^{i{\bar
s}} B_{0,i} {{\bar s} \brack u-i} \alpha(m-i,u-i)$.
\end{proof}

Combining Lemma~\ref{lemma:rk_dual_vector},
Proposition~\ref{prop:trivial_MRD}, and Lemma~\ref{lemma:rk_Asu1},
the rank weight enumerator of $\vspan{{\bf v}}^\perp$ can be
determined at last.

\begin{proposition}\label{prop:rk_W_L}
For ${\bf v} \in \mathrm{GF}(q^m)^n$ with rank $r \geq 0$, the rank
weight enumerator of $\mathcal{L} = \vspan{{\bf v}}^{\perp}$ depends
on only $r$, and is given by
\begin{equation}
    W_\mathcal{L}^{\mbox{\tiny R}}(x,y) = q^{-m} \left\{
    \left[x+(q^m-1)y\right]^{[n]} + (q^m-1) (x-y)^{[r]} *
    \left[x+(q^m-1)y\right]^{[n-r]} \right\}.
\end{equation}
\end{proposition}

\subsection{MacWilliams identity for the rank metric} \label{sec:theorem_rank}

Using the results in Section~\ref{sec:dual_v_rank}, we now derive
the MacWilliams identity for rank metric codes. But first, we give
two definitions from \cite{macwilliams_77} that are needed in our
derivation.

\begin{definition}\label{def:chi}
Let $\mathbb{C}$ be the field of complex numbers. Let $a \in
\mathrm{GF}(q^m)$ and let $\{1,\alpha_1,\ldots,\alpha_{m-1} \}$ be a
basis set of $\mathrm{GF}(q^m)$. We thus have $a = a_0 + a_1
\alpha_1 + \ldots + a_{m-1}\alpha_{m-1}$, where $a_i \in
\mathrm{GF}(q)$ for $0 \leq i \leq m-1$. Finally, let $\zeta \in
\mathbb{C}$ be a primitive $q$-th root of unity, $\chi(a) \df
\zeta^{a_0}$ provides a mapping from $\mathrm{GF}(q^m)$ to
$\mathbb{C}$.
\end{definition}

\begin{definition}[Hadamard transform]\label{def:hadamard} For a mapping $f$ from
$\mathrm{GF}(q^m)^n$ to $\mathbb{C}$, the {\em Hadamard transform}
of $f$, denoted as $\hat{f}$, is defined to be
\begin{equation}\label{eq:hadamard}
    \hat{f}({\bf v}) \df \sum_{{\bf u} \in \mathrm{GF}(q^m)^n} \chi({\bf u} \cdot {\bf
    v}) f({\bf u}),
\end{equation} where ${\bf u} \cdot {\bf
    v}$ denotes the inner product of ${\bf u}$ and ${\bf v}$.
\end{definition}

Let $\mathcal{C}$ be an $(n,k)$ linear code over $\mathrm{GF}(q^m)$,
and let $W_\mathcal{C}^{\mbox{\tiny R}}(x,y) = \sum_{i=0}^n A_i
y^ix^{n-i}$ be its rank weight enumerator and
$W_{\mathcal{C}^{\perp}}^{\mbox{\tiny R}}(x,y) = \sum_{j=0}^n B_j
y^j x^{n-j}$ be the rank weight enumerator of its dual code
$\mathcal{C}^{\perp}$.

\begin{theorem}\label{th:MacWilliams}
For any $(n,k)$ linear code $\mathcal{C}$ and its dual code
$\mathcal{C}^{\perp}$ over $\mathrm{GF}(q^m)$,
\begin{equation}\label{eq:macwilliams}
    W_{\mathcal{C}^{\perp}}^{\mbox{\tiny R}}(x,y) = \frac{1}{|\mathcal{C}|}
    {\bar W}_\mathcal{C}^{\mbox{\tiny R}}\left(x+(q^m-1)y,x-y\right),
\end{equation}
where ${\bar W}_\mathcal{C}^{\mbox{\tiny R}}$ is the $q$-transform
of $W_\mathcal{C}^{\mbox{\tiny R}}$. Equivalently,
\begin{equation}\label{eq:macwilliams2}
    \sum_{j=0}^n B_j y^j x^{n-j} = q^{-mk}\sum_{i=0}^n A_i
    (x-y)^{[i]}* \left[x+(q^m-1)y\right]^{[n-i]}.
\end{equation}
\end{theorem}

\begin{proof}
We have $\rk(\lambda {\bf u}) = \rk({\bf u})$ for all $\lambda \in
\mathrm{GF}(q^m)^*$ and all ${\bf u} \in \mathrm{GF}(q^m)^n$. We
want to determine $\hat{f}_{\mbox{\tiny R}}({\bf v})$ for all ${\bf
v} \in \mathrm{GF}(q^m)^n$. By Definition~\ref{def:hadamard}, we can
split the summation in Eq.~(\ref{eq:hadamard}) into two parts:
$$
    \hat{f}_{\mbox{\tiny R}}({\bf v}) = \sum_{{\bf u} \in \mathcal{L}} \chi({\bf u} \cdot {\bf v})
    f_{\mbox{\tiny R}}({\bf u}) + \sum_{{\bf u} \in F \backslash \mathcal{L}} \chi({\bf u} \cdot
    {\bf v}) f_{\mbox{\tiny R}}({\bf u}),
$$
where $\mathcal{L} = \vspan{{\bf v}}^{\perp}$. If ${\bf u} \in
\mathcal{L}$, then $\chi({\bf u} \cdot {\bf v}) = 1$ by
Definition~\ref{def:chi}, and the first summation is equal to
$W^{\mbox{\tiny R}}_\mathcal{L}(x,y)$. For the second summation, we
gather vectors into groups of the form $\{\lambda {\bf u}_1\}$,
where $\lambda \in \mathrm{GF}(q^m)^*$ and ${\bf u}_1 \cdot {\bf v}
= 1$. We remark that for ${\bf u} \in F \backslash \mathcal{L}$ (see
\cite[Chapter 5, Lemma 9]{macwilliams_77})
$$
    \sum_{\lambda \in \mathrm{GF}(q^m)^*} \chi(\lambda{\bf u}_1 \cdot {\bf
    v})f_{\mbox{\tiny R}}(\lambda {\bf u}_1) =
    f_{\mbox{\tiny R}}({\bf u}_1)\sum_{\lambda \in \mathrm{GF}(q^m)^*} \chi(\lambda) =
    -f_{\mbox{\tiny R}}({\bf u}_1).
$$
Hence the second summation is equal to $-\frac{1}{q^m-1}
W^{\mbox{\tiny R}}_{F\backslash \mathcal{L}}(x,y)$. This leads to
$\hat{f}_{\mbox{\tiny R}}({\bf v}) = \frac{1}{q^m-1} [q^m
W^{\mbox{\tiny R}}_\mathcal{L}(x,y) - W^{\mbox{\tiny R}}_F(x,y)].$
Using $W_F^{\mbox{\tiny R}}(x,y) = [x+(q^m-1)y]^{[n]}$ and
Proposition~\ref{prop:rk_W_L}, we obtain
$\hat{f}_{\mbox{\tiny{R}}}({\bf v}) = (x-y)^{[r]}*
\left[x+(q^m-1)y\right]^{[n-r]}.$

By \cite[Chapter 5, Lemma 11]{macwilliams_77}, any mapping $f$ from
$F$ to $\mathbb{C}$ satisfies $\sum_{{\bf v} \in
\mathcal{C}^{\perp}} f({\bf v}) = \frac{1}{|\mathcal{C}|} \sum_{{\bf
v} \in \mathcal{C}} \hat{f}({\bf v}).$ Applying this result to
$f_{\mbox{\tiny{R}}}({\bf v})$, we obtain (\ref{eq:macwilliams}) and
(\ref{eq:macwilliams2}).
\end{proof}

Also, $B_j$'s can be explicitly expressed in terms of $A_i$'s.

\begin{corollary}\label{cor:krawtchouk}
We have
\begin{equation}\label{eq:B_A_krawtchouk}
    B_j = \frac{1}{|\mathcal{C}|} \sum_{i=0}^n A_i P_j(i;m,n),
\end{equation}
where
\begin{equation}\label{eq:krawtchouk2}
    P_j(i;m,n) \df \sum_{l=0}^j {i \brack l}{n-i \brack j-l}(-1)^l
    q^{\sigma_l}q^{l(n-i)}\alpha(m-l,j-l).
\end{equation}
\end{corollary}

\begin{proof}
We have $(x-y)^{[i]}*(x+(q^m-1)y)^{[n-i]} = \sum_{j=0}^n P_j(i;m,n)
y^jx^{n-j}$. The result follows from Theorem~\ref{th:MacWilliams}.
\end{proof}

Note that although (\ref{eq:B_A_krawtchouk}) is the same as that in
\cite[(3.14)]{delsarte_78}, $P_j(i;m,n)$ in~(\ref{eq:krawtchouk2})
are different from $P_j(i)$ in \cite[(A10)]{delsarte_78} and their
alternative forms in \cite{delsarte_76}. We can show that

\begin{proposition}\label{prop:P=Q}
$P_j(x;m,n)$ in~(\ref{eq:krawtchouk2}) are the generalized
Krawtchouk polynomials.
\end{proposition}

The proof is given in Appendix~\ref{app:prop:P=Q}. Also, it was
pointed out in \cite{delsarte_76} that
$\frac{P_j(x;m,n)}{P_j(0;m,n)}$ is actually a basic hypergeometric
function.

Proposition~\ref{prop:P=Q} shows that $P_j(x;m,n)$
in~(\ref{eq:krawtchouk2}) are an alternative form for $P_j(i)$ in
\cite[(A10)]{delsarte_78}, and hence our results in
Corollary~\ref{cor:krawtchouk} are equivalent to those in
\cite[Theorem~3.3]{delsarte_78}.

\subsection{Moments of the rank distribution}\label{sec:moments_distrib}

Next, we investigate the relationship between moments of the rank
distribution of a linear code and those of its dual code. Our
results parallel those in \cite[p. 131]{macwilliams_77}.

\begin{proposition}\label{prop:bm_x}
For $0 \leq \nu \leq n$,
\begin{equation}\label{eq:bm_x}
     \sum_{i=0}^{n-\nu} {n-i \brack \nu} A_i = q^{m(k-\nu)}
     \sum_{j=0}^{\nu} {n-j \brack n-\nu} B_j.
\end{equation}
\end{proposition}

\begin{proof}
First, apply Eq.~(\ref{eq:B_A_krawtchouk}) to $\mathcal{C}^\perp$.
We obtain $A_i = q^{m(k-n)} \sum_{j=0}^n B_j P_i(j;m,n)$, and hence
$$
    \sum_{i=0}^{n-\nu} {n-i \brack \nu} A_i = q^{m(k-n)}
     \sum_{j=0}^n B_j \sum_{i=0}^n {n-i \brack \nu} P_i(j;m,n).
$$
We have $\sum_{i=0}^n {n-i \brack \nu} P_i(j;m,n) = q^{m(n-\nu)}{n-j
\brack n-\nu}$ \cite[(29)]{delsarte_75}, and the result follows.
\end{proof}

\begin{proof}
First, applying Theorem~\ref{th:MacWilliams} to
$\mathcal{C}^{\perp}$, we obtain
\begin{equation}\label{eq:before_nu}
    \sum_{i=0}^n A_i y^i x^{n-i} = q^{m(k-n)} \sum_{j=0}^n B_j b_j(x,y;m)*a_{n-j}(x,y;m).
\end{equation}

Next, we apply the $q$-derivative with respect to $x$ to
Eq.~(\ref{eq:before_nu}) $\nu$ times. By
Lemma~\ref{lemma:special_q-d} the left hand side (LHS) becomes
$\sum_{i=0}^{n-\nu} \beta(n-i,\nu) A_i y^i x^{n-i-\nu}$, while by
Lemma~\ref{lemma:Leibniz_x} the right hand side (RHS) reduces to
$q^{m(k-n)}\sum_{j=0}^n B_j \psi_j(x,y)$, where
$$
    \psi_j(x,y) \df
    [b_j(x,y;m)*a_{n-j}(x,y;m)]^{(\nu)} =
    \sum_{l=0}^{\nu} {\nu \brack l} q^{(\nu-l)(j-l)} b_j^{(l)}(x,y)*
    a_{n-j}^{(\nu-l)}(x,y;m).
$$
By Lemma~\ref{lemma:special_q-d}, $b_j^{(l)}(x,y;m) = \beta(j,l)
(x-y)^{[j-l]}$ and $a_{n-j}^{(\nu-l)}(x,y;m) =
\beta(n-j,\nu-l)a_{n-j-\nu+l}(x,y;m)$. It can be verified that for
any homogeneous polynomial $b(x,y;m)$ and for any $s \geq 0$,
$(b*a_s)(1,1;m) = q^{ms} b(1,1;m)$. Also, for $x=y=1$,
$b_j^{(l)}(1,1;m) = \beta(j,j) \delta_{j,l}$. We hence have
$\psi_j(1,1) = 0$ for $j>\nu$, and $\psi_j(1,1) = {\nu \brack j}
\beta(j,j) \beta(n-j,\nu-j)q^{m(n-\nu)}$ for $j\leq \nu$. Since
$\beta(n-j,\nu-j) = {n-j \brack \nu-j} \beta(\nu-j,\nu-j)$ and
$\beta(\nu,\nu) = {\nu \brack j} \beta(j,j) \beta(\nu-j,\nu-j)$,
$\psi_j(1,1)= {n-j \brack \nu-j} \beta(\nu,\nu)
q^{m\left(n-\nu\right)}$. Applying $x=y=1$ to the LHS and
rearranging both sides using $\beta(n-i,\nu) = {n-i \brack \nu}
\beta(\nu,\nu)$, we obtain~(\ref{eq:bm_x}).
\end{proof}

Proposition~\ref{prop:bm_x} can be simplified if $\nu$ is less than
the minimum distance of the dual code.

\begin{corollary}\label{cor:binomial_moment_x}
Let $\dr'$ be the minimum rank distance of $\mathcal{C}^{\perp}$. If
$0 \leq \nu < \dr'$, then
\begin{equation}
    \sum_{i=0}^{n-\nu} {n-i \brack \nu} A_i = q^{m(k-\nu)}{n \brack
    \nu}.
\end{equation}
\end{corollary}
\begin{proof}
We have $B_0 = 1$ and $B_1 = \ldots = B_{\nu} = 0$.
\end{proof}

Using the $q^{-1}$-derivative, we obtain another relationship.

\begin{proposition}\label{prop:bm_y}
For $0 \leq \nu \leq n$,
\begin{equation}\label{eq:bm_y}
    \sum_{i=\nu}^n {i \brack \nu} q^{\nu(n-i)} A_i = q^{m(k-\nu)}
    \sum_{j=0}^\nu {n-j \brack n-\nu} (-1)^j q^{\sigma_j}
    \alpha(m-j,\nu-j)q^{j(\nu-j)} B_j.
\end{equation}
\end{proposition}

The proof of Proposition~\ref{prop:bm_y} is similar to that of
Proposition~\ref{prop:bm_x}, and is given in
Appendix~\ref{app:prop:bm_y}. Similarly, when $\nu$ is less than the
minimum distance of the dual code, Proposition~\ref{prop:bm_y} can
be simplified.

\begin{corollary}\label{cor:binomial_moment_y}
If $0 \leq \nu < \dr'$, then
\begin{equation}
    \sum_{i=\nu}^n {i \brack \nu} q^{\nu(n-i)} A_i = q^{m(k-\nu)} {n
    \brack \nu} \alpha(m,\nu).
\end{equation}
\end{corollary}

\begin{proof}
We have $B_0=1$ and $B_1 = \cdots = B_\nu = 0$.
\end{proof}

Following \cite{macwilliams_77}, we refer to the LHS of
Eq.~(\ref{eq:bm_x}) and~(\ref{eq:bm_y}) as moments of the rank
distribution of $\mathcal{C}$.

\subsection{Relation to Delsarte's results}\label{sec:delsarte}


Delsarte \cite{delsarte_78} also derived the MacWilliams identity
for rank metric codes, and below we briefly relate our results to
those by Delsarte.

Delsarte \cite{delsarte_78} considered array codes with the rank
metric. In \cite{delsarte_78}, the inner product between two $m
\times n$ matrices ${\bf A}$ and ${\bf B}$ over $\mathrm{GF}(q)$ is
defined as ${\bf A} \cdot {\bf B} \df \mathrm{Tr}({\bf A} {\bf
B}^T)$. Two matrices ${\bf A}$ and ${\bf B}$ are orthogonal if
$\chi({\bf A} \cdot {\bf B}) = 1$, where $\chi$ is a nontrivial
character of the additive group $\mathrm{GF}(q)$, and dual codes are
defined using this orthogonality. Delsarte then established an
analytical expression (cf. \cite[Theorem~3.3]{delsarte_78}) between
the rank distance enumerator of an array code and that of its dual.

Clearly the definitions of dual codes are different in our work and
\cite{delsarte_78}. However, we show below the two definitions
collide when dual bases are used. With a slight abuse of notation,
the inner products between two vectors and two matrices are both
denoted by $\cdot$ and dual codes of both vector and array codes are
denoted by $\perp$. For all vectors ${\bf x} \in
\mathrm{GF}(q^m)^n$, we expand ${\bf x}$ into a matrix with respect
to the basis $B = \{\beta_i\}_{i=0}^{m-1}$ of $\mathrm{GF}(q^m)$
over $\mathrm{GF}(q)$ and refer to the matrix
$\left\{x_{i,j}\right\}_{i, j=0, 0}^{m-1, n-1}$ as ${\bf x}_B$. That
is, $x_j = \sum_{i=0}^{m-1} x_{i,j} \beta_i$ for $0\leq j < n$. For
a code $\mathcal{C}$ of length $n$ over $\mathrm{GF}(q^m)$, we
denote $\mathcal{C}_B \df \{ {\bf x}_B \in \mathrm{GF}(q)^{m \times
n} | {\bf x} \in \mathcal{C}\}$. Clearly, if $\mathcal{C}$ is an
$(n,k)$ linear code over $\mathrm{GF}(q^m)$, then $\mathcal{C}_B$ is
an $(mn,mk)$ linear array code over $\mathrm{GF}(q)$.

\begin{lemma}\label{lemma:dual_basis}
Let $E = \{\epsilon_i\}_{i=0}^{m-1}$ and $P = \{ \phi_j
\}_{j=0}^{m-1}$ be dual bases of $\mathrm{GF}(q^m)$ over
$\mathrm{GF}(q)$. Then for any ${\bf a}, {\bf b} \in
\mathrm{GF}(q^m)^n$, $\mathrm{Trace}({\bf a} \cdot {\bf b}) = {\bf
a}_E \cdot {\bf b}_P.$
\end{lemma}

\begin{proof}
We have ${\bf a} \cdot {\bf b} = \sum_j a_jb_j = \sum_j \sum_i
\sum_k a_{i,j} \epsilon_i b_{k,j} \phi_k$. Applying the trace
function on both sides, we obtain
$$
    \mathrm{Trace}({\bf a} \cdot {\bf b}) = \sum_j \sum_i \sum_k
    a_{i,j} b_{k,j} \mathrm{Trace}(\epsilon_i \phi_k)
    = \sum_j \sum_i a_{i,j} b_{i,j}
    = \mathrm{Tr}({\bf a}_E {\bf b}_P^T).
$$
\end{proof}

\begin{proposition}\label{prop:equal_duals}
For an $(n,k)$ code $\mathcal{C}$ over $\mathrm{GF}(q^m)$ and dual
bases $E$ and $P$ of $\mathrm{GF}(q^m)$ over $\mathrm{GF}(q)$,
$(\mathcal{C}^\perp)_E = (\mathcal{C}_P)^\perp$.
\end{proposition}

\begin{proof}
Let ${\bf v} \in \mathcal{C}^{\perp}$, then for any ${\bf u} \in
\mathcal{C}$, ${\bf v} \cdot {\bf u} = 0$. Hence ${\bf v}_E \cdot
{\bf u}_P = 0$ by Lemma~\ref{lemma:dual_basis} and $\chi({\bf v}_E
\cdot {\bf u}_P) = 1$ for all ${\bf u}_P \in \mathcal{C}_P$.
Therefore ${\bf v}_E \in (\mathcal{C}_P)^\perp$ and
$(\mathcal{C}^\perp)_E \subseteq (\mathcal{C}_P)^\perp$. Since
$|(\mathcal{C}^\perp)_E| = |(\mathcal{C}_P)^\perp|$, we conclude
that $(\mathcal{C}^\perp)_E = (\mathcal{C}_P)^\perp$.
\end{proof}
Proposition~\ref{prop:equal_duals} implies that our identity in
Corollary~\ref{cor:krawtchouk} can be derived from Delsarte's
identity in \cite[Theorem~3.3]{delsarte_78}. Suppose $\mathcal{C}$
is an $(n,k)$ linear code over $\mathrm{GF}(q^m)$ with rank weight
distribution $A_i$ and its dual code $\mathcal{C}^\perp$ has rank
distribution $B_j$. Let $E$ and $P$ are dual bases of
$\mathrm{GF}(q^m)$ over $\mathrm{GF}(q)$. Note that $\mathcal{C}_P$
and $(\mathcal{C}^\perp)_E$ have the same rank distribution as
$\mathcal{C}$ and $\mathcal{C}^\perp$, respectively. Also by
Proposition~\ref{prop:equal_duals} $\mathcal{C}_P$ and
$(\mathcal{C}^\perp)_E$ are dual array codes. Thus applying
Delsarte's identity to $\mathcal{C}_P$ results in
Corollary~\ref{cor:krawtchouk}. Note that the rank distance
enumerator and rank weight enumerator are the same for linear codes.

Our results in this section are different from Delsarte's results in
several aspects. First, $P_j(i;m,n)$ in~(\ref{eq:krawtchouk2}) are
different from $P_j(i)$ in \cite[(A10)]{delsarte_78} and their
alternative forms in \cite{delsarte_76}. In
Proposition~\ref{prop:P=Q}, we show that $P_j(x;m,n)$ are actually
the generalized Krawtchouk polynomials, and hence $P_j(i;m,n)$
in~(\ref{eq:krawtchouk2}) are equivalent to $P_j(i)$ in
\cite[(A10)]{delsarte_78}. Second, our approach to proving the
MacWilliams identity is different, and intermediate results of our
proof offer interesting insights (see
Lemma~\ref{lemma:rk_dual_vector} and Proposition~\ref{prop:rk_W_L}).
Third, our MacWilliams identity is also expressed in a polynomial
form (Theorem~\ref{th:MacWilliams}) similar to that in
\cite{macwilliams_77}, and the polynomial form allows us to derive
other identities (see, for example, Propositions~\ref{prop:bm_x} and
\ref{prop:bm_y}) that relate the rank distribution of a linear code
to that of its dual.

%

\section{Conclusions}
In this paper, we investigate the packing, covering, and weight
properties of rank metric codes. We show that MRD codes not only are
optimal in the sense of the Singleton bound, but also provide the
optimal solution to the sphere packing problem. We also derive
bounds for the sphere covering problem  and establish the asymptotic
minimum code rate for a code with given relative covering radius.
Finally, we establish identities that relate the rank weight
distribution of a linear code to that of its dual code.

\appendix
The proofs in this section use some well-known properties of
Gaussian polynomials \cite{andrews}:
\begin{eqnarray}
    \label{eq:binomial_symmetry}
    {n \brack k} &=& {n \brack n-k}\\
    \label{eq:binomial_pascal}
    {n \brack k} &=& {n-1 \brack k} + q^{n-k} {n-1 \brack k-1}\\
    \label{eq:binomial_pascal_reversed}
    &=& q^k {n-1 \brack k} + {n-1 \brack k-1}\\
    \label{eq:binomial_n-1}
    &=& \frac{q^n-1}{q^{n-k}-1} {n-1 \brack k}\\
    \label{eq:binomial_k-1}
    &=& \frac{q^{n-k+1}-1}{q^k-1} {n \brack k-1}\\
    \label{eq:binomial_transitive}
    {n \brack k} {k \brack l} &=& {n \brack l} {n-l \brack n-k}.
\end{eqnarray}

\subsection{Proof of Proposition~\ref{prop:excess_bound}}\label{app:prop:excess_bound}
We first establish a key lemma.
\begin{lemma}\label{lemma:A_cap_B}
If ${\bf z} \in Z$ and $0 < \rho < n$, then
\begin{equation}\label{eq:A_cap_B}
    |A \cap B_1({\bf z})| \leq V_1(q^m,n) - q^{\rho-1}{\rho \brack 1}.
\end{equation}
\end{lemma}

\begin{proof}
By definition of $\rho$, there exists ${\bf c} \in C$ such that
$\dr({\bf z},{\bf c}) \leq \rho$. By
Proposition~\ref{prop:inter_2_balls}, $|B_1({\bf z}) \cap
B_{\rho-1}({\bf c})|$ gets its minimal value for $\dr({\bf z},{\bf
c}) = \rho$, which is $q^{\rho-1}{\rho \brack 1}$ by
Proposition~\ref{prop:v-a_problem}. A vector at distance $\leq \rho
- 1$ from any codeword does not belong to $A$. Therefore, $B_1({\bf
z}) \cap B_{\rho-1}({\bf c}) \subseteq B_1({\bf z})\backslash A$,
and hence $|A \cap B_1({\bf z})| = |B_1({\bf z})| - |B_1({\bf z})
\backslash A| \leq V_1(q^m,n) - |B_1({\bf z}) \cap B_{\rho-1}({\bf
c})|$.
\end{proof}

For a code $C$ with covering radius $\rho$ and $\epsilon \geq 1$,
\begin{eqnarray}
    \label{eq:gamma}
    \gamma & \df & \epsilon \left[ q^{mn} - |C|V_{\rho-1}(q^m,n) \right]
    - (\epsilon-1) \left[|C|V_{\rho}(q^m,n) - q^{mn}\right]\\
    \label{eq:vanwee1}
    & \leq & \epsilon |A| - (\epsilon-1)|Z|\\
    \nonumber
    & \leq & \epsilon |A| - (\epsilon-1) |A \cap Z|\\
    \nonumber
    & = & \epsilon |A \backslash Z| + |A \cap Z|,
\end{eqnarray}
where~(\ref{eq:vanwee1}) follows from $|Z| \leq |C|V_{\rho}(q^m,n) -
q^{mn}$, given in Section~\ref{sec:covering_radius}.

\begin{eqnarray}
    \label{eq:vanwee2}
    \gamma & \leq & \sum_{{\bf a} \in A \backslash Z} E_C(B_1({\bf a})) + \sum_{{\bf a} \in A \cap Z}
    E_C(B_1({\bf a}))\\
    \nonumber
    & = & \sum_{{\bf a} \in A} E_C(B_1({\bf a})),
\end{eqnarray}
where~(\ref{eq:vanwee2}) follows from Lemma~\ref{lemma:epsilon} and
$|A \cap Z| \leq E_C(A \cap Z)$.

\begin{eqnarray}
    \label{eq:vanwee3}
    \gamma & \leq & \sum_{{\bf a} \in A} \sum_{{\bf x} \in B_1({\bf a}) \cap
    Z} E_C(\{ {\bf x} \})\\
    \nonumber
    & = & \sum_{{\bf x} \in Z} \sum_{{\bf a} \in B_1({\bf x}) \cap
    A} E_C(\{ {\bf x} \})\\
    \nonumber
    & = & \sum_{{\bf x} \in Z} |A \cap B_1({\bf x})| E_C(\{ {\bf x}
    \}),
\end{eqnarray}
where~(\ref{eq:vanwee3}) follows from the fact the second summation
is over disjoint sets $\{ {\bf x} \}$. Using
Lemma~\ref{lemma:A_cap_B}, we obtain

\begin{eqnarray}
    \nonumber
    \gamma & \leq & \sum_{{\bf x} \in Z} \left(V_1(q^m,n) - q^{\rho-1}{\rho \brack 1}\right) E_C(\{ {\bf x}
    \})\\
    \nonumber
    & = & \left(V_1(q^m,n) - q^{\rho-1}{\rho \brack 1}\right) E_C(Z)\\
    \label{eq:vanwee4}
    & = & \left(V_1(q^m,n) - q^{\rho-1}{\rho \brack 1}\right)(|C|V_{\rho}(q^m,n) - q^{mn}).
\end{eqnarray}

Combining~(\ref{eq:vanwee4}) and~(\ref{eq:gamma}), we
obtain~(\ref{eq:excess_bound}).

\subsection{Proof of Proposition~\ref{prop:bound_K_12.1}}
\label{app:prop:bound_K_12.1}

\begin{proof}
Given a radius $\rho$ and a code $C$, denote by $P_\rho(C)$ the set
of vectors in $\mathrm{GF}(q^m)^n$ that are at distance $> \rho$
from $C$. To simplify notations, $Q \df q^{mn}$ and $p_\rho(C) \df
Q^{-1} |P_\rho(C)|$. Let us denote the set of all codes over
$\mathrm{GF}(q^m)$ of length $n$ and cardinality $K$ as $S_K$.
Clearly $|S_K| = {Q \choose K}$. Let us calculate the average value
of $p_\rho(C)$ for all codes $C \in S_K$:
\begin{eqnarray}
    \nonumber
    \frac{1}{|S_K|} \sum_{C \in S_K} p_\rho(C) &=& \frac{1}{|S_K|} Q^{-1} \sum_{C \in S_K} |P_\rho(C)|
    = \frac{1}{|S_K|} Q^{-1} \sum_{C \in S_K} \sum_{{\bf x} \in F | \dr({\bf x}, C) > \rho}
    1\\
    \nonumber
    &=& \frac{1}{|S_K|} Q^{-1} \sum_{{\bf x} \in F} \sum_{C \in S_K| \dr({\bf x}, C) >
    \rho} 1\\
    \label{eq:Q_choose_K}
    &=& \frac{1}{|S_K|} Q^{-1} \sum_{{\bf x} \in F} {Q - V_\rho(q^m,n) \choose K}\\
    &=& {Q - V_\rho(q^m,n) \choose K} \left/ {Q \choose K} \right. .
\end{eqnarray}
Eq.~(\ref{eq:Q_choose_K}) comes from the fact that there are ${Q -
V_\rho(q^m,n) \choose K}$ codes with cardinality $K$ that do not
cover ${\bf x}$. For all $K$, there exists a code $C' \in S_K$ for
which $p_\rho(C')$ is no more than the average, that is:
\begin{eqnarray}
    \nonumber
    p_\rho(C') 
    &\leq& {Q \choose K}^{-1} {Q - V_\rho(q^m,n) \choose K}\\
    \nonumber
    &\leq& \left( 1-Q^{-1}V_\rho(q^m,n)\right)^K.
\end{eqnarray}

Let us choose $K = \left\lfloor  -\frac{1}{\log_Q
\left(1-Q^{-1}V_\rho(q^m,n) \right)} \right\rfloor + 1$ so that $K
\log_Q \left(1 - Q^{-1} V_\rho(q^m,n) \right) < -1$ and hence
$p_\rho(C') = \left(1- Q^{-1} V_\rho(q^m,n) \right)^K < Q^{-1}$. It
follows that $|P_\rho(C')| < 1$, and $C'$ has covering radius at
most $\rho$.
\end{proof}

\subsection{Proof of Lemma~\ref{lemma:special_prod}}
\label{app:lemma:special_prod} The  proof is by induction on $l$.
Clearly all the claims hold for $l=0$. Suppose $y^{[{\bar l}-1]} =
q^{\sigma_{{\bar l} - 1}}y^{{\bar l} - 1}$ for ${\bar l}\geq 1$,
then $y^{[{\bar l}]} = y^{[{\bar l}-1]}*y = q^{{\bar
l}-1}q^{\sigma_{{\bar l}-1}}y^{\bar l} = q^{\sigma_{\bar l}}y^{\bar
l}$. The proof for $x^{[l]}$ is similar.

Suppose Eq.~(\ref{eq:x+y^s}) holds for $l = {\bar l}-1$. We have
$a_{\bar l}(x,y;m)= a_{{\bar l}-1}(x,y;m)* (x+(q^m-1)y)=
\sum_{u=0}^l a_{{\bar l},u} y^u x^{{\bar l}-u}$. By
(\ref{eq:q-product}), we have
\begin{eqnarray}
    \nonumber
    a_{{\bar l},u} &=& q^u a_{{\bar l}-1,u} + q^{u-1}a_{{\bar l}-1,u-1}(q^{m-u+1}-1)\\
    \nonumber
    &=& q^u {{\bar l}-1 \brack u} \alpha(m,u) + q^{u-1} {{\bar l}-1 \brack u-1}
    \alpha(m,u-1) (q^{m-u+1}-1)\\
    \nonumber
    &=& \left(q^u{{\bar l}-1 \brack u} + {{\bar l}-1 \brack u-1}\right) \alpha(m,u)\\
    \label{eq:a}
    &=& {{\bar l} \brack u}\alpha(m,u),
\end{eqnarray}
where Eq.~(\ref{eq:a}) follows Eq.~(\ref{eq:binomial_pascal}).

Suppose Eq.~(\ref{eq:x-y^r}) holds for $l = {\bar l}-1$. We have
$b_{\bar l}(x,y;m)= b_{{\bar l}-1}(x,y;m)* (x-y)= \sum_{u=0}^l
b_{{\bar l},u} y^u x^{{\bar l}-u}$. By (\ref{eq:q-product}), we have
\begin{eqnarray}
    b_{{\bar l},u} &=& q^u b_{{\bar l}-1,u} - q^{u-1} b_{{\bar l}-1,u-1}\nonumber \\
    &=& q^u {{\bar l}-1 \brack u}(-1)^u q^{\sigma_u} + q^{u-1} {{\bar l}-1 \brack
    u-1}(-1)^uq^{\sigma_{u-1}}\nonumber \\
    &=& \left(q^u{{\bar l}-1 \brack u} + {{\bar l}-1 \brack u-1}\right)(-1)^u q^{\sigma_u}\nonumber \\
    \label{eq:b'}
    &=& {{\bar l} \brack u}(-1)^u q^{\sigma_u},
\end{eqnarray}
where Eq.~(\ref{eq:b'}) also follows Eq.~(\ref{eq:binomial_pascal}).

\subsection{Proof of Lemma~\ref{lemma:special_q-d}}
\label{app:lemma:special_q-d} The proof is by induction on $\nu$.
Clearly all the claims hold for $\nu = 0$. The $\nu$-th
$q$-derivative of $x^l$ follows Definition~\ref{def:q-derivative}.
Suppose Eq.~(\ref{eq:al_nu}) holds for $\nu = {\bar \nu}-1$, then
\begin{eqnarray}
a^{({\bar \nu})}_l(x,y;m) &=& \left[ \beta\left(l,\bar{\nu}-1\right)
a_{l-\bar{\nu}+1}(x,y; m) \right]^{(1)} \nonumber \\
&=&
\beta(l,{\bar \nu}-1) \sum_{i=1}^{l-{\bar \nu}+1}
    {l-{\bar \nu}+1 \brack i} \alpha(m,l-{\bar \nu}+1-i)
    y^{l-{\bar \nu}+1-i} {i \brack 1} x^{i-1}\nonumber \\
    \label{eq:e}
    &=& \beta(l,{\bar \nu}-1) {l-{\bar \nu}+1 \brack 1} \sum_{i=0}^{l-{\bar \nu}}
    {l-{\bar \nu} \brack i} \alpha(m,l-{\bar \nu} - i)
    y^{l-{\bar \nu}-i}x^{i}\\
    \nonumber
    &=& \beta(l,{\bar \nu}) a_{l-{\bar \nu}}(x,y;m),
\end{eqnarray}
where Eq.~(\ref{eq:e}) follows Eq.~(\ref{eq:binomial_transitive}).

Similarly, suppose Eq.~(\ref{eq:bl_nu}) holds for $\nu = {\bar
\nu}-1$, hence
\begin{eqnarray*}
    b^{({\bar \nu})}_l(x,y;m) &=& \left[ \beta\left(l,\bar{\nu}-1\right) b_{l-\bar{\nu}+1}(x,y; m) \right]^{(1)} \nonumber \\
    &=& \beta(l,{\bar \nu} -1) \sum_{i=1}^{l-{\bar \nu}+1}
    {l-{\bar \nu}+1 \brack i} (-1)^{l-{\bar \nu}+1-i} q^{\sigma_{l-{\bar \nu}+1-i}}
    y^{l-{\bar \nu}+1-i} {i \brack 1} x^{i-1}\\
    &=& \beta(l,{\bar \nu}) \sum_{i=0}^{l-{\bar \nu}} {l-{\bar \nu}
    \brack i} (-1)^{l-{\bar \nu}-i} q^{\sigma_{l-{\bar \nu}-i}}
    y^{l-{\bar \nu}-i}x^i\\
    &=& \beta(l,{\bar \nu}) b_{l-{\bar \nu}}(x,y;m).
\end{eqnarray*}

\subsection{Proof of Lemma~\ref{lemma:Leibniz_x}}
\label{app:lemma:leibniz_x}

We consider homogeneous polynomials $f(x,y;m) = \sum_{i=0}^r f_i y^i
x^{r-i}$ and $u(x,y;m) = \sum_{i=0}^r u_i y^i x^{r-i}$ of degree $r$
as well as $g(x,y;m) = \sum_{j=0}^s g_j y^j x^{s-j}$ and $v(x,y;m) =
\sum_{j=0}^s v_j y^j x^{s-j}$ of degree $s$. First, we need a
technical lemma.

\begin{lemma}\label{lemma:u*v}
If $u_r = 0$, then
\begin{equation}\label{eq:ux}
    \frac{1}{x}(u(x,y;m)*v(x,y;m)) = \frac{u(x,y;m)}{x}*v(x,y;m).
\end{equation}

If $v_s = 0$, then
\begin{equation}\label{eq:vx}
    \frac{1}{x}(u(x,y;m)*v(x,y;m)) = u(x,qy;m)*\frac{v(x,y;m)}{x}.
\end{equation}
\end{lemma}

\begin{proof}
Suppose $u_r=0$, then $\frac{u(x,y;m)}{x} = \sum_{i=0}^{r-1} u_i y^i
x^{r-1-i}$. Hence
\begin{eqnarray*}
    \frac{u(x,y;m)}{x}*v(x,y;m) &=& \sum_{k=0}^{r+s-1}
    \left( \sum_{l=0}^k q^{ls} u_l(m) v_{k-l}(m-l) \right)
    y^k x^{r+s-1-k}\\
    &=& \frac{1}{x}(u(x,y;m)*v(x,y;m)).
\end{eqnarray*}

Suppose $v_s=0$, then $\frac{v(x,y;m)}{x} = \sum_{j=0}^{s-1} v_j y^j
x^{s-1-j}$. Hence
\begin{eqnarray*}
    u(x,qy;m)*\frac{v(x,y;m)}{x} &=& \sum_{k=0}^{r+s-1}
    \left( \sum_{l=0}^k q^{l(s-1)} q^l u_l(m) v_{k-l}(m-l) \right)
    y^k x^{r+s-1-k}\\
    &=& \frac{1}{x}(u(x,y;m)*v(x,y;m)).
\end{eqnarray*}
\end{proof}

We now give a proof of Lemma~\ref{lemma:Leibniz_x}.

\begin{proof}
In order to simplify notations, we omit the dependence of the
polynomials $f$ and $g$ on the parameter $m$. The proof is by
induction on $\nu$.
For $\nu=1$, we have
\begin{eqnarray}
    \nonumber
    [f(x,y)*g(x,y)]^{(1)} &=& \frac{1}{(q-1)x} \big[f(qx,y)*g(qx,y) - f(qx,y)*g(x,y) \cdots\\
    \nonumber
    &+& f(qx,y)*g(x,y) - f(x,y)*g(x,y) \big]\\
    \nonumber
    &=& \frac{1}{(q-1)x}\left[ f(qx,y)*(g(qx,y)-g(x,y)) +
    (f(qx,y)-f(x,y))*g(x,y)\right]\\
    \label{eq:x_b}
    &=& f(qx,qy)*\frac{g(qx,y)-g(x,y)}{(q-1)x} +
    \frac{f(qx,y)-f(x,y)}{(q-1)x}*g(x,y)\\
    \label{eq:x_b1}
    &=& q^r f(x,y)*g^{(1)}(x,y) + f^{(1)}(x,y)*g(x,y),
\end{eqnarray}
where Eq.~(\ref{eq:x_b}) follows Lemma~\ref{lemma:u*v}.

Now suppose Eq.~(\ref{eq:leibniz_nu}) is true for $\nu = {\bar
\nu}$. In order to further simplify notations, we omit the
dependence of the various polynomials in $x$ and $y$. We have
\begin{eqnarray}
    \nonumber
    (f*g)^{({\bar \nu}+1)} &=& \sum_{l=0}^{\bar \nu} {{\bar \nu} \brack l}
    q^{({\bar \nu}-l)(r-l)} \left[ f^{(l)}*g^{({\bar \nu}-l)}\right]^{(1)}\\
    \label{eq:x_c}
    &=& \sum_{l=0}^{{\bar \nu}} {{\bar \nu} \brack l}
    q^{({\bar \nu}-l)(r-l)} \left( q^{r-l}f^{(l)}*g^{({\bar \nu}-l+1)} +
    f^{(l+1)}*g^{({\bar \nu}-l)} \right)\\
    \nonumber
    &=& \sum_{l=0}^{{\bar \nu}} {{\bar \nu} \brack l}
    q^{({\bar \nu}+1-l)(r-l)} f^{(l)}*g^{({\bar \nu}-l+1)} + \sum_{l=1}^{{\bar \nu}+1} {{\bar \nu} \brack l-1}
    q^{({\bar \nu}+1-l)(r-l+1)} f^{(l)}*g^{({\bar \nu}-l+1)}\\
    \nonumber
    &=& \sum_{l=1}^{{\bar \nu}} \left( {{\bar \nu} \brack l} + q^{{\bar \nu}+1-l} {{\bar \nu}
    \brack l-1} \right) q^{({\bar \nu}+1-l)(r-l)} f^{(l)}*g^{({\bar \nu}-l+1)}
    + q^{({\bar \nu}+1)r}f*g^{({\bar \nu}+1)} + f^{({\bar \nu}+1)}*g\\
    \label{eq:x_d}
    &=& \sum_{l=0}^{{\bar \nu}+1} {{\bar \nu}+1 \brack l} q^{({\bar \nu}+1-l)(r-l)}
    f^{(l)}*g^{({\bar \nu}-l+1)},
\end{eqnarray}
where Eq.~(\ref{eq:x_c}) follows Eq.~(\ref{eq:x_b1}), and
Eq.~(\ref{eq:x_d}) follows Eq.~(\ref{eq:binomial_pascal}).
\end{proof}

\subsection{Proof of Lemma~\ref{lemma:special_q-1_d}} \label{app:lemma:special_q-1_d}

\begin{proof}
The proof is by induction on $\nu$. Clearly all the claims hold for
$\nu=0$. The $\nu$-th $q^{-1}$-derivative of $y^l$ follows from
Definition~\ref{def:q-1_derivative}. Suppose
Eq.~(\ref{eq:al_nu_q-1}) holds for $\nu = {\bar \nu}-1$, then

\begin{eqnarray*}
    a_l^{ \{ {\bar \nu} \} }(x,y;m) &=& \beta(l,{\bar \nu}-1) q^{\sigma_{\bar \nu-1}}
    \alpha(m,{\bar \nu}-1) a_{l-{\bar \nu}+1}^{\{1 \}} (x,y;m)\\
   &=& \beta(l,{\bar \nu}-1) q^{\sigma_{{\bar \nu}-1}}
    \alpha(m,{\bar \nu}-1) \sum_{i=0}^{l-{\bar \nu}+1} {l-{\bar \nu}+1 \brack i}
    \alpha(m-{\bar \nu}+1,i) x^{l-{\bar \nu}+1-i} q^{1-i} {i \brack 1} y^{i-1}\\
    &=& \beta(l,{\bar \nu}) q^{\sigma_{{\bar \nu}-1}} \alpha(m,{\bar \nu}-1)
    \sum_{i=1}^{l-\bar \nu+1} {l-\bar \nu \brack i-1}
    (q^{m-\bar \nu+1} - 1) \alpha(m-\bar \nu,i-1) x^{l-\bar \nu+1-i} y^{i-1}\\
    &=& \beta(l,\bar \nu) q^{\sigma_{\bar \nu}} \alpha(m,\bar \nu)
    a_{l-\bar \nu}(x,y;m-\bar \nu).
\end{eqnarray*}

Suppose Eq.~(\ref{eq:bl_nu_q-1}) holds for $\nu = {\bar \nu} - 1$,
then

\begin{eqnarray*}
    b_l^{ \{{\bar \nu}\} }(x,y;m) &=& (-1)^{{\bar \nu}-1}
    \beta(l,{\bar \nu}-1) b_{l-{\bar \nu}+1}^{\{ 1\}}(x,y;m)\\
    &=& (-1)^{\bar \nu-1} \beta(l,\bar \nu)
    \sum_{i=1}^{l-\bar \nu+1} {l-\bar \nu+1 \brack i-1} (-1)^i q^{\sigma_i}
    x^{l-\bar \nu+1-i} q^{1-i} y^{i-1}\\
    &=& (-1)^{\bar \nu}\beta(l,\bar \nu)b_{l-\bar \nu}(x,y;m).
\end{eqnarray*}
\end{proof}

\subsection{Proof of Lemma~\ref{lemma:Leibniz_y}}
\label{app:lemma:leibniz_y}

We consider homogeneous polynomials $f(x,y;m) = \sum_{i=0}^r f_i y^i
x^{r-i}$ and $u(x,y;m) = \sum_{i=0}^r u_i y^i x^{r-i}$ of degree $r$
as well as $g(x,y;m) = \sum_{j=0}^s g_j y^j x^{s-j}$ and $v(x,y;m) =
\sum_{j=0}^s v_j y^j x^{s-j}$ of degree $s$. First, we need a
technical lemma.

\begin{lemma}\label{lemma:u*v_y}
If $u_0 = 0$, then
\begin{equation}\label{eq:uy}
    \frac{1}{y}(u(x,y;m))*v(x,y;m)) = q^s \frac{u(x,y;m)}{y}*v(x,y;m-1).
\end{equation}

If $v_0 = 0$, then
\begin{equation}\label{eq:vy}
    \frac{1}{y}(u(x,y;m)*v(x,y;m)) = u(x,qy;m)*\frac{v(x,y;m)}{y}.
\end{equation}
\end{lemma}

\begin{proof}
Suppose $u_0=0$, then $\frac{u(x,y;m)}{y} = \sum_{i=0}^{r-1} u_{i+1}
x^{r-1-i} y^i$. Hence
\begin{eqnarray*}
    q^s \frac{u(x,y;m)}{y}*v(x,y;m-1) &=& q^s \sum_{k=0}^{r+s-1}
    \left( \sum_{l=0}^k q^{ls} u_{l+1} v_{k-l}(m-1-l) \right)
    x^{r+s-1-k}y^k\\
    &=& q^s \sum_{k=1}^{r+s}
    \left( \sum_{l=1}^k q^{(l-1)s} u_l v_{k-l}(m-l) \right)
    x^{r+s-k}y^{k-1}\\
    &=& \frac{1}{y}(u(x,y;m)*v(x,y;m)).
\end{eqnarray*}

Suppose $v_0=0$, then $\frac{v(x,y;m)}{y} = \sum_{j=0}^{s-1} v_{j+1}
x^{s-1-j} y^j$. Hence
\begin{eqnarray*}
    u(x,qy;m)*\frac{v(x,y;m)}{y} &=& \sum_{k=0}^{r+s-1}
    \left( \sum_{l=0}^k q^{l(s-1)} q^l u_l v_{k-l+1}(m-l) \right)
    x^{r+s-1-k}y^k\\
    &=& \sum_{k=1}^{r+s}
    \left( \sum_{l=0}^{k-1} q^{ls} u_l v_{k-l}(m-l) \right)
    x^{r+s-k}y^{k-1}\\
    &=& \frac{1}{y}(u(x,y;m)*v(x,y;m)).
\end{eqnarray*}
\end{proof}

We now give a proof of Lemma~\ref{lemma:Leibniz_y}.

\begin{proof}
The proof is by induction on $\nu$. For $\nu=0$, the result is
trivial. For $\nu=1$, we have
\begin{eqnarray}
    \nonumber
    [f(x,y;m)*g(x,y;m)]^{\{ 1 \}} &=& \frac{1}{(q^{-1}-1)y} \big[ f(x,q^{-1}y;m) * g(x,q^{-1}y;m)
    - f(x,q^{-1}y;m)*g(x,y;m)\\
    \nonumber
    &+& f(x,q^{-1}y;m)*g(x,y;m) - f(x,y;m)*g(x,y;m) \big]\\
    \nonumber
    &=& \frac{1}{(q^{-1}-1)y} \big[ f(x,q^{-1}y;m)*(g(x,q^{-1}y;m)-g(x,y;m)) \\
    \nonumber
    &+& (f(x,q^{-1}y;m)-f(x,y;m))*g(x,y;m) \big]\\
    \nonumber
    &=& f(x,y;m)*\frac{g(x,q^{-1}y;m) - g(x,y;m)}{(q^{-1}-1)y}\\
    \label{eq:y_b}
    &+& q^s\frac{f(x,q^{-1}y;m) - f(x,y;m)}{(q^{-1}-1)y} * g(x,y;m-1)\\
    \label{eq:y_c}
    &=& f(x,y;m)*g^{\{1\}}(x,y;m) + q^s f^{\{1\}}(x,y;m)*g(x,y;m-1),
\end{eqnarray}
where~(\ref{eq:y_b}) comes from Lemma~\ref{lemma:u*v_y}.

Now suppose Equation~(\ref{eq:leibniz_y_nu}) is true for ${\bar
\nu}$. In order to further simplify notations, we omit the
dependence of the various polynomials in $x$ and $y$. We have
\begin{eqnarray}
    \nonumber
    [f(m)*g(m)]^{\{ {\bar \nu}+1 \}} &=& \sum_{l=0}^{\bar \nu} {{\bar \nu} \brack l}
    q^{l(s-{\bar \nu}+l)} \left[ f^{\{ l \}}(m)*g^{\{{\bar \nu}-l\}}(m-l)\right]^{\{1\}}\\
    \nonumber
    &=& \sum_{l=0}^{{\bar \nu}} {{\bar \nu} \brack l}
    q^{l(s-{\bar \nu}+l)} \Big( f^{\{ l \}}(m)*g^{\{{\bar
    \nu}-l+1\}}(m-l)\\
    \label{eq:y_d}
    &+& q^{s-{\bar \nu}+l} f^{\{l+1\}}(m)*g^{\{{\bar \nu}-l\}}(m-l-1) \Big)\\
    \nonumber
    &=& \sum_{l=0}^{{\bar \nu}} {{\bar \nu} \brack l}
    q^{l(s-{\bar \nu}+l)} f^{\{ l \}}(m)*g^{\{{\bar \nu}-l+1\}}(m-l) \\
    \nonumber
    &+& \sum_{l=1}^{{\bar \nu}+1} {{\bar \nu} \brack l-1}
    q^{l(s-{\bar \nu}+l-1)} f^{\{ l \}}(m)*g^{\{{\bar \nu}-l+1\}}(m-l)\\
    \nonumber
    &=& \sum_{l=1}^{{\bar \nu}} \left( q^l {{\bar \nu} \brack l} + {{\bar \nu}
    \brack l-1} \right) q^{l(s-{\bar \nu}+l-1)}
    f^{\{ l \}}(m)*g^{\{{\bar \nu}-l+1\}}(m-l)\\
    \nonumber
    &+& f(m)*g^{\{{\bar \nu}+1\}}(m) + q^{s({\bar \nu}+1)}f^{\{{\bar \nu}+1\}}(m)*g(m-{\bar \nu}-1)\\
    \label{eq:y_e}
    &=& \sum_{l=0}^{{\bar \nu}+1} {{\bar \nu}+1 \brack l} q^{l(s-{\bar \nu}-1+l)}
    f^{\{ l \}}(m)*g^{\{{\bar \nu}-l+1\}}(m-l),
\end{eqnarray}
where~(\ref{eq:y_d}) comes from~(\ref{eq:y_c}) and~(\ref{eq:y_e})
comes from~(\ref{eq:binomial_pascal_reversed}).
\end{proof}

\subsection{Proof of Proposition~\ref{prop:P=Q}}\label{app:prop:P=Q}

\begin{proof}
It was shown in \cite{delsarte_76} that the generalized Krawtchouk
polynomials are the only solutions to the recurrence
\begin{equation}\label{eq:recurrence_Pji}
    P_{j+1}(i+1;m+1,n+1) = q^{j+1} P_{j+1}(i+1;m,n) + q^j
    P_j(i;m,n)
\end{equation}
with initial conditions $P_j(0;m,n) = {n \brack j} \alpha(m,j)$.
Clearly, our polynomials satisfy these initial conditions. We hence
show that $P_j(i;m,n)$ satisfy the recurrence in
Eq.~(\ref{eq:recurrence_Pji}). We have
\begin{eqnarray}
    \nonumber
    P_{j+1}(i+1;m+1,n+1) &=& \sum_{l=0}^{i+1} {i+1 \brack l} {n-i \brack
    j+1-l} (-1)^l q^{\sigma_l} q^{l(n-i)} \alpha(m+1-l,j+1-l)\\
    \nonumber
    &=& \sum_{l=0}^{i+1} {i+1 \brack l} {m+1-l \brack
    j+1-l} (-1)^l q^{\sigma_l} q^{l(n-i)} \alpha(n-i,j+1-l)\\
    &=& \sum_{l=0}^{i+1} \left\{ q^l{i \brack l} + {i \brack l-1} \right\}
    \left\{ q^{j+1-l} {m-l \brack j+1-l} + {m-l \brack j-l} \right\}
    \cdots \nonumber \\
    &\cdots& (-1)^l q^{\sigma_l} q^{l(n-i)} \alpha(n-i,j+1-l) \label{eq:P=Q_a} \\
    \nonumber
    &=& \sum_{l=0}^i {i \brack l}
    q^{j+1} {m-l \brack j+1-l} (-1)^l q^{\sigma_l} q^{l(n-i)} \alpha(n-i,j+1-l)\\
    \nonumber
    &+& \sum_{l=0}^i q^l {i \brack l} q^{j+1} {m-l \brack j-l}
    (-1)^l q^{\sigma_l} q^{l(n-i)} \alpha(n-i,j+1-l)\\
    \nonumber
    &+& \sum_{l=1}^{i+1} {i \brack l-1}
    q^{j+1-l} {m-l \brack j+1-l} (-1)^l q^{\sigma_l} q^{l(n-i)} \alpha(n-i,j+1-l)\\
    \label{eq:P=Q1}
    &+& \sum_{l=1}^{i+1} {i \brack l-1} {m-l \brack j-l}
    (-1)^l q^{\sigma_l} q^{l(n-i)} \alpha(n-i,j+1-l),
\end{eqnarray}
where~(\ref{eq:P=Q_a}) follows
from~(\ref{eq:binomial_pascal_reversed}). Let us denote the four
summations in the right hand side of Eq.~(\ref{eq:P=Q1}) as $A$,
$B$, $C$, and $D$ respectively. We have $A = q^{j+1}
P_{j+1}(i;m,n)$, and
\begin{eqnarray}
    \label{eq:P=Q_b}
    B &=& \sum_{l=0}^i {i \brack l} {m-l \brack j-l}
    (-1)^l q^{\sigma_l} q^{l(n-i)} \alpha(n-i,j-l) (q^{n-i+l}-q^j),\\
    \nonumber
    C &=& \sum_{l=0}^{i} {i \brack l}
    q^{j-l} {m-l-1 \brack j-l} (-1)^{l+1} q^{\sigma_{l+1}} q^{(l+1)(n-i)}
    \alpha(n-i,j-l)\\
    \label{eq:P=Q_c}
    &=& -q^{j+n-i} \sum_{l=0}^{i} {i \brack l}
    {m-l \brack j-l} (-1)^l q^{\sigma_l} q^{l(n-i)}\alpha(n-i,j-l)
    \frac{q^{m-j}-1}{q^{m-l}-1},\\
    \label{eq:P=Q_d}
    D &=& -q^{n-i} \sum_{l=0}^i {i \brack l} {m-l \brack j-l}
    (-1)^l q^{\sigma_{l+1}} q^{l(n-i)} \alpha(n-i,j-l) q^l
    \frac{q^{j-l}-1}{q^{m-l}-1},
\end{eqnarray}
where~(\ref{eq:P=Q_c}) follows from~(\ref{eq:binomial_n-1})
and~(\ref{eq:P=Q_d}) follows from both~(\ref{eq:binomial_k-1})
and~(\ref{eq:binomial_n-1}).
Combining~(\ref{eq:P=Q_b}),~(\ref{eq:P=Q_c}), and~(\ref{eq:P=Q_d}),
we obtain
\begin{eqnarray*}
    B+C+D &=& \sum_{l=0}^i {i \brack l} {m-l \brack j-l}
    (-1)^l q^{\sigma_{l+1}} q^{l(n-i)} \alpha(n-i,j-l) \cdots\\
    &\cdots& \left\{ q^{n-i+l}-q^j - q^{n-i}\frac{q^{m}-q^j}{q^{m-l}-1}
    -q^{n-i} \frac{q^{j}-q^l}{q^{m-l}-1}\right\}\\
    &=& -q^j P_j(i;m,n).
\end{eqnarray*}

\end{proof}

\subsection{Proof of Proposition~\ref{prop:bm_y}}
\label{app:prop:bm_y}

Before proving Proposition~\ref{prop:bm_y}, we need two technical
lemmas.

\begin{lemma}\label{lemma:delta}
For all $m$, $\nu$, and $l$, we have
\begin{equation}\label{eq:delta}
    \delta(m,\nu,j) \df \sum_{i=0}^j {j \brack i} (-1)^i
    q^{\sigma_i} \alpha(m-i,\nu) = \alpha(\nu,j) \alpha(m-j,\nu-j) q^{j(m-j)}.
\end{equation}
\end{lemma}

\begin{proof}
The proof is by induction on $j$. When $j=0$, the claim trivially
holds. Let us suppose it holds for ${\bar j}$. We have
\begin{eqnarray}
    \nonumber
    \delta(m,\nu,{\bar j}+1) &=& \sum_{i=0}^{{\bar j}+1} {{\bar j}+1 \brack i} (-1)^i
    q^{\sigma_i} \alpha(m-i,\nu)\\
    \label{eq:delta_a}
    &=& \sum_{i=0}^{{\bar j}+1} \left( q^i {{\bar j} \brack i} + {{\bar j} \brack i-1} \right) (-1)^i
    q^{\sigma_i} \alpha(m-i,\nu)\\
    \nonumber
    &=& \sum_{i=0}^{\bar j} q^i {{\bar j} \brack i} (-1)^i q^{\sigma_i}
    \alpha(m-i,\nu) + \sum_{i=1}^{{\bar j}+1} {{\bar j} \brack i-1} (-1)^i
    q^{\sigma_i} \alpha(m-i,\nu)\\
    \nonumber
    &=& \sum_{i=0}^{\bar j} q^i {{\bar j} \brack i} (-1)^i q^{\sigma_i}
    \alpha(m-i,\nu) - \sum_{i=0}^{{\bar j}} {{\bar j} \brack i} (-1)^i
    q^{\sigma_{i+1}} \alpha(m-1-i,\nu)\\
    \nonumber
    &=& \sum_{i=0}^{\bar j} q^i {{\bar j} \brack i} (-1)^i q^{\sigma_i}
    \alpha(m-1-i,\nu-1)q^{m-1-i} (q^\nu - 1)\\
    \nonumber
    &=& q^{m-1}(q^\nu - 1) \delta(m-1,\nu-1,{\bar j})\\
    \nonumber
    &=& \alpha(\nu,{\bar j}+1) \alpha(m-{\bar j}-1,\nu-{\bar j}-1) q^{({\bar j}+1)(m-{\bar
    j}-1)},
\end{eqnarray}
where~(\ref{eq:delta_a}) follows
from~(\ref{eq:binomial_pascal_reversed}).
\end{proof}

\begin{lemma}\label{lemma:theta}
For all $n$, $\nu$, and $j$, we have
\begin{equation}\label{eq:theta}
    \theta(n,\nu,j) \df \sum_{l=0}^j {j \brack l}
    {n-j \brack \nu-l} q^{l(n-\nu)} (-1)^l q^{\sigma_l}
    \alpha(\nu-l,j-l) = (-1)^j q^{\sigma_j} {n-j \brack n-\nu}.
\end{equation}
\end{lemma}

\begin{proof}
The proof goes by induction on $j$. When $j=0$, the claim trivially
holds. Let us suppose it holds for ${\bar j}$. We have
\begin{eqnarray}
    \nonumber
    \theta(n,\nu,{\bar j}+1) &=& \sum_{l=0}^{{\bar j}+1} {{\bar j}+1 \brack l}
    {n-1-{\bar j} \brack \nu-l} q^{l(n-\nu)} (-1)^l q^{\sigma_l}
    \alpha(\nu-l,{\bar j}+1-l)\\
    \label{eq:theta1}
    &=& \sum_{l=0}^{{\bar j}+1} \left( {{\bar j} \brack l} + q^{{\bar j}+1-l} {{\bar j} \brack l-1}
    \right) {n-1-{\bar j} \brack \nu-l} q^{l(n-\nu)} (-1)^l q^{\sigma_l}
    \alpha(\nu-l,{\bar j}+1-l)\\
    \nonumber
    &=& \sum_{l=0}^{\bar j} {{\bar j} \brack l} {n-1-{\bar j} \brack \nu-l} q^{l(n-\nu)} (-1)^l q^{\sigma_l}
    \alpha(\nu-l,{\bar j}-l) (q^{\nu-l} - q^{{\bar j}-l})\\
    \label{eq:theta2}
    &+& \sum_{l=1}^{{\bar j}+1} q^{{\bar j}-l+1} {{\bar j} \brack l-1} {n-1-{\bar j} \brack \nu-l} q^{l(n-\nu)} (-1)^l q^{\sigma_l}
    \alpha(\nu-l,{\bar j}-l+1),
\end{eqnarray}
where~(\ref{eq:theta1}) follows from~(\ref{eq:binomial_pascal}). Let
us denote the first and second summations in the right hand side of
(\ref{eq:theta2}) as $A$ and $B$, respectively. We have
\begin{eqnarray}
    \nonumber
    A &=& (q^\nu - q^{\bar j}) \sum_{l=0}^{\bar j} {{\bar j} \brack l} {n-1-{\bar j} \brack \nu-l} q^{l(n-1-\nu)} (-1)^l q^{\sigma_l}
    \alpha(\nu-l,{\bar j}-l)\\
    \nonumber
    &=& (q^\nu - q^{\bar j}) \theta(n-1,\nu,{\bar j})\\
    \label{eq:theta_A}
    &=& (q^\nu-q^{\bar j}) (-1)^{\bar j} q^{\sigma_{\bar j}} {n-1-{\bar j} \brack n-1-\nu},
\end{eqnarray}
and
\begin{eqnarray}
    \nonumber
    B &=& \sum_{l=0}^{\bar j} q^{{\bar j}-l} {{\bar j} \brack l} {n-1-{\bar j} \brack \nu-1-l} q^{(l+1)(n-\nu)} (-1)^{l+1} q^{\sigma_{l+1}}
    \alpha(\nu-1-l,{\bar j}-l)\\
    \nonumber
    &=& - q^{{\bar j}+n-\nu} \sum_{l=0}^{\bar j} {{\bar j} \brack l} {n-1-{\bar j} \brack \nu-1-l} q^{l(n-\nu)} (-1)^l q^{\sigma_l}
    \alpha(\nu-1-l,{\bar j}-l)\\
    \nonumber
    &=& -q^{{\bar j}+n-\nu} \theta(n-1,\nu-1,{\bar j})\\
    \label{eq:theta_B}
    &=& -q^{{\bar j}+n-\nu} (-1)^{\bar j} q^{\sigma_{\bar j}} {n-1-{\bar j} \brack n-\nu}.
\end{eqnarray}

Combining~(\ref{eq:theta1}), (\ref{eq:theta_A}),
and~(\ref{eq:theta_B}), we obtain
\begin{eqnarray}
    \nonumber
    \theta(n,\nu,{\bar j}+1) &=& (-1)^{\bar j} q^{\sigma_{\bar j}}
    \left\{ (q^\nu-q^{\bar j}){n-1-{\bar j} \brack n-1-\nu} - q^{{\bar j}+n-\nu} {n-1-{\bar j} \brack n-\nu} \right\}\\
    \label{eq:theta3}
    &=& (-1)^{{\bar j}+1} q^{\sigma_{{\bar j}+1}} {n-1-{\bar j} \brack n-\nu}
    \left\{ - (q^{\nu-{\bar j}}-1) \frac{q^{n-\nu}-1}{q^{\nu-{\bar j}} - 1} + q^{n-\nu}
    \right\}\\
    &=& (-1)^{{\bar j}+1} q^{\sigma_{{\bar j}+1}} {n-1-{\bar j} \brack
    n-\nu},
\end{eqnarray}
where~(\ref{eq:theta3}) follows from~(\ref{eq:binomial_k-1}).
\end{proof}

We now give a proof of Proposition~\ref{prop:bm_y}.

\begin{proof}
We apply the $q^{-1}$-derivative with respect to $y$ to
Eq.~(\ref{eq:before_nu}) $\nu$ times, and we apply $x=y=1$. By
Lemma~\ref{lemma:special_q-1_d} the left hand side (LHS) becomes
\begin{equation}
    \sum_{i=\nu}^n q^{\nu(1-i) + \sigma_\nu} \beta(i,\nu) A_i
    = q^{\nu(1-n) + \sigma_\nu} \beta(\nu,\nu) \sum_{i=\nu}^n
    {i \brack \nu} q^{\nu(n-i)} A_i.
\end{equation}
The right hand side (RHS) becomes $q^{m(k-n)} \sum_{j=0}^n B_j
\psi_j(1,1)$, where
\begin{eqnarray}
    \nonumber
    \psi_j(x,y) &\df& \left[ b_j(x,y;m) * a_{n-j}(x,y;m)
    \right]^{ \{\nu\}}\\
    \label{eq:psi1}
    &=& \sum_{l=0}^\nu {\nu \brack l} q^{l(n-j-\nu+l)}
    b_j^{\{l\}}(x,y;m) * a_{n-j}^{\{ \nu-l \}}(x,y;m-l)\\
    \label{eq:psi2}
    &=& \sum_{l=0}^\nu {\nu \brack l} q^{l(n-j-\nu+l)}
    (-1)^l \beta(j,l) \beta(n-j,\nu-l) q^{-\sigma_{\nu-l}}\cdots\\
    \nonumber
    &\cdots& b_{j-l}(x,y;m) * \alpha(m-l,\nu-l)
    a_{n-j-\nu+l}(x,y;m-\nu)\\
    \nonumber
    &=& \beta(\nu,\nu) q^{-\sigma_\nu} \sum_{l=0}^\nu {j \brack l}
    {n-j \brack \nu-l} q^{l(n-j)} (-1)^l q^{\sigma_l}\cdots\\
    \nonumber
    &\cdots& b_{j-l}(x,y;m) * \alpha(m-l,\nu-l)
    a_{n-j-\nu+l}(x,y;m-\nu),
\end{eqnarray}
where~(\ref{eq:psi1}) and~(\ref{eq:psi2}) follow from
Lemmas~\ref{lemma:Leibniz_y} and~\ref{lemma:special_q-1_d}
respectively.

We have
\begin{eqnarray}
    \nonumber
    && \left[ b_{j-l} * \alpha(m-l,\nu-l) a_{n-j-\nu+l} \right] (1,1;m-\nu) \cdots\\
    \nonumber
    &=& \sum_{u=0}^{n-\nu} \left[ \sum_{i=0}^u q^{i(n-j-\nu+l)}
    {j-l \brack i} (-1)^i q^{\sigma_i} \alpha(m-i-l,\nu-l) {n-j-\nu+l
    \brack u-i} \alpha(m-\nu-i,u-i) \right]\\
    \nonumber
    &=& q^{(m-\nu)(n-\nu-j+l)} \sum_{i=0}^{j-l} {j-l \brack i}
    (-1)^i q^{\sigma_i} \alpha(m-l-i,\nu-l)\\
    \label{eq:psi_a}
    &=& q^{(m-\nu)(n-\nu-j+l)} \alpha(\nu-l,j-l)
    \alpha(m-j,\nu-j)q^{(j-l)(m-j)},
\end{eqnarray}
where~(\ref{eq:psi_a}) follows from Lemma~\ref{lemma:delta}. Hence
\begin{eqnarray}
    \nonumber
    \psi_j(1,1) &=& \beta(\nu,\nu) q^{m(n-\nu) + \nu(1-n) + \sigma_\nu}
    \alpha(m-j,\nu-j)q^{j(\nu-j)}\cdots\\
    \nonumber
    &\cdots&\sum_{l=0}^j {j \brack l}
    {n-j \brack \nu-l} q^{l(n-\nu)} (-1)^l q^{\sigma_l}
    \alpha(\nu-l,j-l)\\
    \label{eq:psi_b}
    &=& \beta(\nu,\nu) q^{m(n-\nu) + \nu(1-n) + \sigma_\nu}
    \alpha(m-j,\nu-j)q^{j(\nu-j)} (-1)^j q^{\sigma_j} {n-j \brack
    n-\nu},
\end{eqnarray}
where~(\ref{eq:psi_b}) follows from Lemma~\ref{lemma:theta}.
Incorporating this expression for $\psi_j(1,1)$ in the definition of
the RHS and rearranging both sides, we obtain the result.
\end{proof}

\bibliographystyle{IEEETran}
\bibliography{gpt}

\begin{thebibliography}{10}
\providecommand{\url}[1]{#1}
\csname url@rmstyle\endcsname
\providecommand{\newblock}{\relax}
\providecommand{\bibinfo}[2]{#2}
\providecommand\BIBentrySTDinterwordspacing{\spaceskip=0pt\relax}
\providecommand\BIBentryALTinterwordstretchfactor{4}
\providecommand\BIBentryALTinterwordspacing{\spaceskip=\fontdimen2\font plus
\BIBentryALTinterwordstretchfactor\fontdimen3\font minus
  \fontdimen4\font\relax}
\providecommand\BIBforeignlanguage[2]{{%
\expandafter\ifx\csname l@#1\endcsname\relax
\typeout{** WARNING: IEEEtran.bst: No hyphenation pattern has been}%
\typeout{** loaded for the language `#1'. Using the pattern for}%
\typeout{** the default language instead.}%
\else
\language=\csname l@#1\endcsname
\fi
#2}}

\bibitem{Hua51}
L.~Hua, ``A theorem on matrices over a field and its applications,''
  \emph{Chinese Mathematical Society}, vol.~1, no.~2, pp. 109--163, 1951.

\bibitem{delsarte_78}
P.~Delsarte, ``Bilinear forms over a finite field, with applications to coding
  theory,'' \emph{Journal of Combinatorial Theory A}, vol.~25, pp. 226--241,
  1978.

\bibitem{tarokh_98}
V.~Tarokh, N.~Seshadri, and A.~R. Calderbank, ``Space-time codes for high data
  rate wireless communication: Performance criterion and code construction,''
  \emph{IEEE Trans. Info. Theory}, vol.~44, pp. 774--765, March 1998.

\bibitem{lusina_it03}
P.~Lusina, E.~M. Gabidulin, and M.~Bossert, ``Maximum rank distance codes as
  space-time codes,'' \emph{IEEE Trans. Info. Theory}, vol.~49, pp. 2757--2760,
  Oct. 2003.

\bibitem{gabidulin_lncs91}
E.~M. Gabidulin, A.~V. Paramonov, and O.~V. Tretjakov, ``Ideals over a
  non-commutative ring and their application in cryptology,'' \emph{LNCS}, vol.
  573, pp. 482--489, 1991.

\bibitem{gabidulin_pit0285}
E.~M. Gabidulin, ``Optimal codes correcting lattice-pattern errors,''
  \emph{Problems on Information Transmission}, vol.~21, no.~2, pp. 3--11, 1985.

\bibitem{roth_it91}
R.~M. Roth, ``Maximum-rank array codes and their application to crisscross
  error correction,'' \emph{IEEE Trans. Info. Theory}, vol.~37, no.~2, pp.
  328--336, March 1991.

\bibitem{gabidulin_pit0185}
E.~M. Gabidulin, ``Theory of codes with maximum rank distance,'' \emph{Problems
  on Information Transmission}, vol.~21, no.~1, pp. 1--12, Jan. 1985.

\bibitem{babu_95}
N.~{Suresh Babu}, ``Studies on rank distance codes,'' Ph.D Dissertation, IIT
  Madras, Feb. 1995.

\bibitem{chen_mn96}
K.~Chen, ``On the non-existence of perfect codes with rank distance,''
  \emph{Mathematische Nachrichten}, vol. 182, pp. 89--98, 1996.

\bibitem{roth_it97}
R.~M. Roth, ``Probabilistic crisscross error correction,'' \emph{IEEE Trans.
  Info. Theory}, vol.~43, no.~5, pp. 1425--1438, Sept. 1997.

\bibitem{vasantha_gs99}
W.~B. Vasantha and N.~{Suresh Babu}, ``On the covering radius of rank-distance
  codes,'' \emph{Ganita Sandesh}, vol.~13, pp. 43--48, 1999.

\bibitem{kshevetskiy_isit05}
A.~Kshevetskiy and E.~M. Gabidulin, ``The new construction of rank codes,''
  \emph{Proc. IEEE Int. Symp. on Information Theory}, pp. 2105--2108, Sept.
  2005.

\bibitem{gabidulin_isit05}
E.~M. Gabidulin and P.~Loidreau, ``On subcodes of codes in the rank metric,''
  \emph{Proc. IEEE Int. Symp. on Information Theory}, pp. 121--123, Sept. 2005.

\bibitem{gadouleau_globecom06}
M.~Gadouleau and Z.~Yan, ``Properties of codes with the rank metric,''
  \emph{Proceedings of 2006 IEEE Globecom}, pp. 1--5, November 2006.

\bibitem{gadouleau_itw06}
------, ``Decoder error probability of {MRD} codes,'' \emph{Proceedings of IEEE
  International Theory Workshop}, pp. 264--268, October 2006.

\bibitem{gadouleau_it06}
------, ``Error performance analysis of maximum rank distance codes,''
  \emph{Submitted to IEEE Transactions on Information Theory}, available at
  http://arxiv.org/pdf/cs.IT/0612051.

\bibitem{loidreau_07}
P.~Loidreau, ``Properties of codes in rank metric,'' available at
  http://arxiv.org/pdf/cs.DM/0610057.

\bibitem{SE05}
{M. Schwartz and T. Etzion}, ``{Two-dimensional cluster-correcting codes},''
  \emph{IEEE Trans. Info. Theory}, vol.~51, no.~6, pp. 2121--2132, June 2005.

\bibitem{gabidulin_it07}
E.~M. Gabidulin and P.~Loidreau, ``Properties of subspace subcodes of optimum
  codes in rank metric,'' available at http://arxiv.org/pdf/cs.IT/0607108.

\bibitem{loidreau_05}
P.~Loidreau, ``A {W}elch-{B}erlekamp like algorithm for decoding {G}abidulin
  codes,'' \emph{Proceedings of the 4th International Workshop on Coding and
  Cryptography}, 2005.

\bibitem{richter_isit04}
G.~Richter and S.~Plass, ``Fast decoding of rank-codes with rank errors and
  column erasures,'' \emph{Proceedings of IEEE ISIT 2004}, p. 398, June 2004.

\bibitem{delsarte_73}
P.~Delsarte, ``Four fundamental parameters of a code and their combinatorial
  significance,'' \emph{Information and Control}, vol.~23, pp. 407--438, 1973.

\bibitem{berger_book71}
T.~Berger, \emph{Rate Distortion Theory: A Mathematical Basis for Data
  Compression}, ser. Information and System Sciences Series, T.~Kailath,
  Ed.\hskip 1em plus 0.5em minus 0.4em\relax Englewood Cliffs, N.J.:
  Prentice-Hall, 1971.

\bibitem{macwilliams_77}
F.~MacWilliams and N.~Sloane, \emph{The Theory of Error-Correcting
  Codes}.\hskip 1em plus 0.5em minus 0.4em\relax Amsterdam: North-Holland,
  1977.

\bibitem{blahut_83}
R.~Blahut, \emph{Theory and Practice of Error Control Codes}.\hskip 1em plus
  0.5em minus 0.4em\relax Addison-Wesley, 1983.

\bibitem{cohen_book97}
G.~D. Cohen, I.~Honkala, S.~Litsyn, and A.~C. Lobstein, \emph{Covering
  Codes}.\hskip 1em plus 0.5em minus 0.4em\relax Elsevier, 1997.

\bibitem{vasantha_itw1002}
W.~B. Vasantha and R.~J. Selvaraj, ``Multi-covering radii of codes with rank
  metric,'' \emph{Proc. Information Theory Workshop}, p. 215, Oct. 2002.

\bibitem{loidreau_01}
P.~Loidreau, ``{\'E}tude et optimisation de cryptosyst\`emes \`a cl\'e publique
  fond\'es sur la th\'eorie des codes correcteurs,'' Ph.D. Dissertation,
  \'Ecole Polytechnique, Paris, France, May 2001.

\bibitem{vanwee_88}
G.~van Wee, ``Improved sphere bounds on the covering radius of codes,''
  \emph{IEEE Trans. Info. Theory}, vol.~34, pp. 237--245, 1988.

\bibitem{vanwee_91}
------, ``Bounds on packings and coverings by spheres in $q$-ary and mixed
  {H}amming spaces,'' \emph{Journal of Combinatorial Theory, Series A},
  vol.~57, pp. 116--129, 1991.

\bibitem{andrews}
G.~E. Andrews, \emph{The Theory of Partitions}, ser. Encyclopedia of
  Mathematics and its Applications, G.-C. Rota, Ed.\hskip 1em plus 0.5em minus
  0.4em\relax Reading, MA: Addison-Wesley, 1976, vol.~2.

\bibitem{kleitman_66}
D.~J. Kleitman, ``On a combinatorial conjecture of {E}rd\"os,'' \emph{Journal
  of Combinatorial Theory}, vol.~1, pp. 209--214, 1966.

\bibitem{cohen_it86}
G.~D. Cohen, A.~C. Lobstein, and N.~J.~A. Sloane, ``Further results on the
  covering radius of codes,'' \emph{IEEE Trans. Info. Theory}, vol.~32, pp.
  680--694, 1986.

\bibitem{gasper_book04}
G.~Gasper and M.~Rahman, \emph{Basic Hypergeometric Series}, 2nd~ed., ser.
  Encyclopedia of Mathematics and its Applications.\hskip 1em plus 0.5em minus
  0.4em\relax Cambridge University Press, 2004, vol.~96.

\bibitem{berger_it03}
T.~Berger, ``Isometries for rank distance and permutation group of {G}abidulin
  codes,'' \emph{IEEE Trans. Info. Theory}, vol.~49, no.~11, pp. 3016--3019,
  November 2003.

\bibitem{delsarte_76}
P.~Delsarte, ``Properties and applications of the recurrence $f(i+1,k+1,n+1) =
  q^{k+1} f(i,k+1,n) - q^k f(i,k,n)$,'' \emph{SIAM Journal of Applied
  Mathematics}, vol.~31, no.~2, pp. 262--270, September 1976.

\bibitem{delsarte_75}
P.~Delsarte and J.~Goethals, ``Alternating bilinear forms over
  $\mathrm{GF}(q)$,'' \emph{Journal of Combinatorial Theory}, vol.~19, pp.
  26--50, 1975.

\end{thebibliography}

\begin{table}
\begin{center}
\begin{tabular}{|c|c|cccccc|}
    \hline
    $m$ & $n$ & $\rho=1$ & $\rho=2$ & $\rho=3$ & $\rho=4$ &
    $\rho=5$ & $\rho=6$\\
    \hline
    2 & 2 & b 3-4 A & 1 & & & &\\
    \hline
    3 & 2 & b 4 B & 1 & & & &\\
      & 3 & b 11-32 C & a 2-4 C & 1 & & &\\
    \hline
    4 & 2 & b 7-8 B & 1 & & & &\\
      & 3 & b 40-64 B & b 4-8 C & 1 & & &\\
      & 4 & c 293-1024 C & b 10-64 C & a 2-8 C & 1 & &\\
    \hline
    5 & 2 & b 12-16 B & 1 & & & &\\
      & 3 & b 154-256 B & b 6-8 B & 1 & & &\\
      & 4 & b 2267-4096 B & b 33-256 C & a 3-8 C & 1 & &\\
      & 5 & b 34894-$2^{17}$ C & b 233-2979 E & b 10-128 C & a 2-8 C & 1&\\
    \hline
    6 & 2 & b 23-32 B & 1 & & & &\\
      & 3 & b 601-1024 B & a 10-16 B & 1 & &  &\\
      & 4 & b 17822-$2^{15}$ B & b 123-256 B & b 6-16 C & 1 &  &\\
      & 5 & b 550395-$2^{20}$ B & b 1770-$2^{14}$ C & c 31-256 C & a 3-16 C & 1 &\\
      & 6 & c 17318410-$2^{26}$ C & c 27065-424990 E & c 214-4299 E & c 9-181 D &
      a 2-16 C & 1\\
    \hline
    7 & 2 & b 44-64 B & 1 & & & &\\
      & 3 & b 2372-4096 B & a 19-32 B & 1 & & & \\
      & 4 & b 141231-$2^{18}$ B & c 484-1024 B & b 10-16 B & 1 & & \\
      & 5 & b 8735289-$2^{24}$ B & b 13835-$2^{15}$ B & b 112-1024 C & a 5-16 C & 1 &\\
      & 6 & b 549829402-$2^{30}$ B & c 42229-$2^{22}$ C & b 1584-$2^{15}$ C & b 31-746 E &
      a 3-16 C & 1\\
      & 7 & b 34901004402-$2^{37}$ C & c 13205450-244855533 E & b 23978-596534 E & c 203-5890 E &
      a 8-242 D & a 2-16 C\\
    \hline
\end{tabular}
\caption{Bounds on $\Kr(q^m,n,\rho)$, for $2 \leq m \leq 7$, $2 \leq
n \leq m$, and $1 \leq \rho \leq 6$. For each set of parameters, the
tightest lower and upper bounds on $\Kr(q^m,n,\rho)$ are given, and
letters associated with the numbers are used to indicate the
tightest bound. The lower case letters \textup{a--c} correspond to
the lower bounds
in~(\ref{eq:obvious_bounds_K}),~(\ref{eq:bound_K_cohen}),
and~(\ref{eq:excess_bound}) respectively. The upper case letters
A--E denote the upper bounds
in~(\ref{eq:obvious_bounds_K}),~(\ref{eq:bound_K_MRD1}),~(\ref{eq:bound_K_mixed}),~(\ref{eq:bound_K_12.1}),
and~(\ref{eq:bound_JSL}) respectively.}\label{table:bounds}
\end{center}
\end{table}

\begin{table}[!htp]
\begin{center}
\begin{tabular}{|c|c|ccccc|}
    \hline
    $m$ & $n$ & $\rho=2$ & $\rho=3$ & $\rho=4$ & $\rho=5$ &
    $\rho=6$\\
    \hline
    4 & 4 & 1-2 & 1 & 0 & & \\
    \hline
    5 & 4 & 1-2 & 1 & 0 & & \\
      & 5 & 2-3 & 1-2 & 1 & 0 & \\
    \hline
    6 & 4 & 2 & 1 & 0 & & \\
      & 5 & 2-3 & 1-2 & 1 & 0 & \\
      & 6 & 3-4 & 2-3 & 1-2 & 1 & 0\\
    \hline
    7 & 4 & 2 & 1 & 0 & & \\
      & 5 & 2-3 & 1-2 & 1 & 0 & \\
      & 6 & 3-4 & 2-3 & 1-2 & 1 & 0\\
      & 7 & 4-5 & 3-4 & 2-3 & 1-2 & 1\\
    \hline
    8 & 4 & 2 & 1 & 0 & & \\
      & 5 & 3 & 2 & 1 & 0 & \\
      & 6 & 3-4 & 2-3 & 1-2 & 1 & 0\\
      & 7 & 4-5 & 3-4 & 2-3 & 1-2 & 1\\
      & 8 & 5-6 & 3-5 & 2-4 & 1-3 & 1-2\\
    \hline
\end{tabular}
\caption{Bounds on $k$ for $q=2$, $4 \leq m \leq 8$, $4 \leq n \leq
m$, and $2 \leq \rho \leq 6$.}\label{table:linear_bounds}
\end{center}
\end{table}

\end{document}